\documentstyle[eqsecnum,aps,prb,preprint]{revtex}
\tightenlines

\begin{document}
%\draft
%\preprint{ Tex version}
%--------------------------------------------------------------------------
\title{
       Electronic structure and magnetism of \\
       Fe$_{3-x}$V$_{x}$X  (X = Si, Ga and Al) alloys
       by the KKR-CPA  method
}
\author{ A. Bansil }
\address{
         Department of Physics,
         Northeastern University, Boston, Massachusetts 02115  }
\author{S. Kaprzyk }
\address{
         Department of Physics,
         Northeastern University, Boston, Massachusetts 02115, \\
         and \\
         Faculty of Physics and Nuclear Techniques,
         Academy of Mining and  Metallurgy,         \\
         al. Mickiewicza 30, 30-073 Krak\'ow, Poland  }
\author{ P.E. Mijnarends }
\address{
         Interfaculty Reactor Institute, Delft University of Technology, \\
         Mekelweg 15, 2629 JB Delft, The Netherlands,  \\
         and \\
         Department of Physics,
         Northeastern University, Boston, Massachusetts 02115 }
\author{ J. Tobo{\l}a }
\address{
         Faculty of Physics and Nuclear Techniques,
         Academy of Mining and  Metallurgy,         \\
         al. Mickiewicza 30, 30-073 Krak\'ow, Poland  }
\date{\today}
\maketitle
%--------------------------------------------------------------------------
\begin{abstract}

We present first principles charge- and spin-selfconsistent
electronic structure computations on the Heusler-type disordered alloys
Fe$_{3-x}$V$_{x}$X for three different metalloids X=(Si, Ga and Al).
In these calculations we use the methodology based on the
Korringa-Kohn-Rostoker formalism and the coherent-potential approximation
(KKR-CPA), generalized to treat disorder in multi-component complex
alloys. Exchange correlation effects are incorporated within
the local spin density (LSD) approximation.
Total energy calculations for Fe$_{3-x}$V$_{x}$Si show that V
substitutes preferentially on the Fe(B) site, not on the Fe(A,C) site, in
agreement with experiment. Furthermore,
calculations have been carried out for Fe$_{3-x}$V$_{x}$X alloys
(with, $x$ = 0.25, 0.50 and 0.75), together with the end compounds
Fe$_{3}$X and Fe$_{2}$VX, and the limiting cases of a single V impurity
in Fe$_{3}$X and a single Fe(B) impurity in Fe$_{2}$VX.
We delineate clearly how the electronic states
and magnetic moments at various sites in Fe$_{3-x}$V$_{x}$X
evolve as a function of the V content and the
metalloid valence. Notably, the spectrum
of Fe$_{3-x}$V$_{x}$X (X=Al and Ga) develops
a pseudo-gap for the majority as well as minority
spin states around the Fermi energy in the V-rich regime
which, together with local moments of Fe(B) impurities,
may play a role in the anomalous behavior of the transport properties.
The total magnetic moment in Fe$_{3-x}$V$_{x}$Si is found to
decrease {\em non-linearly}, and the Fe(B)
moment to {\em increase} with increasing $x$;
this is in contrast to expectations of
the `local environment' model, which holds that
the total moment should vary linearly while the Fe(B) moment
should remain constant. The common-band model which describes
the formation of bonding and antibonding states with different
weights on the different atoms, however, provides insight into
the electronic structure of this class of compounds.

\end{abstract}


\narrowtext

% -------------------------------------------------------------------

\section{ Introduction}
\label{sec:Intr}

Heusler-type ternary and pseudo-binary compounds\cite{heusler03}
Y$_2$ZX in the L2$_{1}$ or DO$_{3}$ structure,
where Y and Z denote metal atoms and X is a metalloid,
display a remarkably rich variety of behavior
in their electronic, magnetic and transport properties.
Among early studies of the Fe-based alloys, a note may be
made of the work on
FeAl,\cite{fallot36,arrot59,oles65}
FeSi,\cite{foex38,benoit55} Fe$_3$Al,\cite{bradley32,nathans58}
Fe$_3$Si,\cite{stearns68,moss72}
and some compounds.\cite{webster71,ziebeck74,ziebeck76}
Niculescu {\em et al.}\cite{niculescu83} give an extensive 
review of Fe$_{3-x}$T$_{x}$Si alloys
for various transition metals T.
The electronic structure of the Heusler compounds can range
from       metallic to semimetallic or semiconducting.
A number of cases of half-metallic ferromagnetic phases,
where the system is metallic for one spin direction and
semiconducting for the other,\cite{groot83,groot86}
have been identified. Examples of these are
Co$_{2-x}$Fe$_{x}$MnSi\cite{fujii90,tobola92} and
Co$_{2}$MnSi$_{1-x}$Ge$_{x}$,\cite{ishida95}
and the existence of anti-ferromagnetic ordering in some
instances has been discussed.\cite{leuken95,endo95}
It is often possible to substitute on a specific metal
site in the lattice with other magnetic or non-magnetic atoms,
thereby inducing continuous changes in physical characteristics.
\cite{pierre94,kouacou95,tobola96}
In view of their tunable magnetic and transport properties,
these compounds have attracted wide attention as potential
electronic materials suitable for various applications.
\cite{kronmuller96}
With all this in mind, it is hardly surprising that Heusler-type
compounds have been the subject of numerous theoretical
and experimental studies over the years.

Recently, the compound Fe$_2$VAl has attracted special attention
because of the intriguing behavior of its electrical resistivity,
specific heat, and photoelectric properties.\cite{nishino97} The
resistivity
shows semiconductorlike behavior with a negative temperature
coefficient suggesting an energy gap of $\sim 0.1$ eV. The
photoelectron spectrum on the other hand seems to show a 
Fermi edge which precludes the existence of an energy gap wider
than a few hundredths of an eV. Finally, low-temperature specific
heat measurements for $T \rightarrow 0$ yield a term $\gamma T$
with $\gamma \sim 14$ mJ/mol K$^2$, which results in an effective
mass about 20 times as large as the bare electron band mass. This
mass enhancement is thought to originate from spin fluctuations
\cite{singh98,guo98} or from excitonic correlations.\cite{weht98}
It makes Fe$_2$VAl a possible candidate for a 3d heavy-fermion
system. A similar resistivity behavior has been reported  for
Fe$_{3-x}$V$_x$Ga with $x$ near 1.0,\cite{kawamiya91} while
Fe$_{2.4}$V$_{0.6}$Si shows an onset of the same behavior.\cite{nishino93}

Concerning relevant {\em theoretical} studies, there is a substantial
body of literature devoted to work on a variety of ordered
Heusler-type phases and related systems. Among systems of
present interest we mention
Fe$_{3}$Si,\cite{switendick76,garba86}
Fe$_{3}$Ga,\cite{ishida89}
Fe$_{3}$Al,\cite{ishida76,haydock80,williams82} and
Fe$_2$VAl.\cite{singh98,guo98,weht98}           Little has been
done on the disordered phases; we are only aware of the
study of Fe$_{3-x}$V$_{x}$Si alloys in
Ref.~\onlinecite{kudrnovsky91}, which is based on a
non-selfconsistent crystal potential.

In the present article we report extensive first principles
electronic structure computations on Fe$_{3-x}$V$_{x}$X over the
entire composition range for three different metalloids X,
namely, Si, Ga, and Al. We consider the parent systems Fe$_3$X
for which $x=0$, while setting $x=1$ yields the corresponding compounds
Fe$_2$VX. By choosing X=Ga or X=Al the effect of replacing Si by either
trivalent Ga or Al has been studied, and by
varying $x$ the entire composition range of disordered alloys is
covered.

It is well-known that Heusler-type compounds display remarkable
`site-selectivity' properties\cite{burch81,burch82}
in the sense that substituted metal atoms
show a preference for entering the lattice in specific
crystallographic positions.
In the generic compound Y$_{3-x}$Z$_{x}$X,
the metal atom Z generally prefers B sites in the lattice
if Z lies to the left of Y in the periodic table,
and the A or C positions if Z lies to the right of Y.
We have carried out total energy calculations which confirm
this trend in Fe$_{3-x}$V$_x$Si.

An issue of interest concerns the validity of the
so-called `local environmental' model
(reviewed in Ref. \onlinecite{niculescu83})
which holds that the Fe moment scales with the number of Fe atoms
in the nearest neighbor (nn) shell. The dilution of
Fe(B) with non-magnetic V atoms in Fe$_3$Si will then
yield a linear decrease in the Fe(A,C) moment, while the Fe(B)
moment remains unchanged. The reason is that V substitution does not
change the nn-shell of Fe(B) which continues to contain 8 Fe(A,C) atoms,
while the number of Fe atoms in the nn-shell of Fe(A,C) decreases
progressively.
Our computations show substantial deviations from this simple
picture in Fe$_{3-x}$V$_{x}$X, and imply that interactions
beyond the nn-shell play a significant role in the behavior of
electronic structure and magnetic moments.
The computed spectra also yield insight into a number
of other issues, such as the applicability of a rigid-band type
picture in describing the effects of substitutions
in Fe$_3$Si,\cite{switendick76}
and the extent to which the ferromagnetism in Fe$_{3-x}$V$_{x}$X
\cite{fujii94} can be modeled as a rigid splitting of the paramagnetic bands.
Finally, we clarify the nature of carriers in Fe$_{3-x}$V$_{x}$X
as a function of composition with consequences for transport phenomena
\cite{kawamiya91,nishino93,kawamiya82}
in these materials.

We have attempted to make contact with relevant experiments as far
as possible. Although our primary interest is in the composition dependence
of various physical quantities, some intercomparisons
for the end compounds Fe$_3$X and Fe$_2$VX are undertaken.
The specific experimental data considered are:
(i) the composition dependence of the total magnetic moment in
             Fe$_{3-x}$V$_{x}$X for the three different metalloids,
(ii) the site specific magnetic moments in the end compounds
             Fe$_3$X and Fe$_2$VX,
and (iii) the soft x-ray emission spectra
             of Jia {\em et al.}\cite{jia92} on Fe$_3$Si.
Our theoretical predictions concerning the detailed variation of
magnetic moments on the Fe(B), Fe(A,C), V, and Si sites in
Fe$_{3-x}$V$_{x}$X show interesting trends which should prove
worthwhile to investigate experimentally.

Finally, a few words about our theoretical methods are in order.
    The disorder is treated within the framework of the charge- and
spin-selfconsistent KKR-CPA methodology which we have developed
and implemented
in order to handle multi-component random alloys in a highly robust
manner.\cite{tobola92,kaprzyk90,bansil91,foota,ban87,ban82,he76,sto84}
We use the  generalized  tetrahedron method\cite{kaprzyk86}
to carry out {\bf k} space integrations in disordered muffin-tin alloys,
and as a result, our KKR-CPA codes allow us to treat the
end compounds (Fe$_3$X and Fe$_2$VX) as well as the
properties of single impurities in these limiting cases within
a consistent, unified theoretical framework.
We have also generalized the Lloyd formula\cite{kaprzyk90}
for the total number of states below any energy
to multi-component alloys in an analytically
satisfactory manner, permitting us an accurate evaluation of the
Fermi energy in all cases.
We are thus able to delineate clearly for the first time
how the majority and minority spin states and magnetic moments
in Fe$_3$Si develop when the Fe(B) position in the lattice is
substituted by V atoms, and/or when Si is replaced by a metalloid of
different valence. The results presented here are highly accurate
and involve no parameters other than the
experimental lattice constants, and
constitute a reliable basis for testing the
underlying KKR-CPA and local spin density (LSD) approximations.

An outline of this article is as follows. The introductory remarks
are followed in Section~\ref{sec:Over} by an overview
of our KKR-CPA formalism for multi-component alloys.
The specific formulae used in computing various physical
quantities discussed in this article are stated.
Section~\ref{sec:Stru} outlines the relevant structural aspects of the
Fe$_{3-x}$V$_{x}$X compounds. Section~\ref{sec:Comp} summarizes
some technical details involved in our computations.
Section~\ref{sec:Resu} takes up the discussion and presentation
of the results, and is divided into a number of subsections
in view of the quantity and complexity of the material involved.
An effort has been made to keep the presentation as brief as possible.

%%%
% ----------------end section Introduction---------------------------

\section{ An Overview of the spin-dependent KKR-CPA formalism
          for multi-component complex alloys }
\label{sec:Over}

We consider a multi-component complex alloy where the Bravais lattice
is  defined by lattice vectors ${\bf R}_n$ $\{n=1,...,N\}$ with basis
atoms in positions ${\bf a}_k$  $\{k=1,...,K\}$. $K$-sublattices
can be generated from the basis points ${\bf a}_k$ via lattice
translations ${\bf R}_n$. For simplicity, we assume that one of
these  sublattices, $k_{CP}$, is occupied randomly by two
types of atoms, A and B, with concentrations $c_A$ and $c_B$,
respectively. Other sublattices are taken to be perfectly ordered.
\cite{footnote3}
The KKR-CPA formalism of interest here proceeds within the framework
of an effective one electron Hamiltonian
\cite{tobola92,foota,ban87,ban82,he76,sto84,bansil93,kaprzyk97}
                                         where the crystal potential
is assumed to possess the form of non-overlapping muffin-tin
spheres of radius $S_k$, i.e., the potential is spherically
symmetric around each atom and constant (usually defined as the
potential zero) in the interstitial region.
Although such a Hamiltonian is more appropriate
for a close packed metallic system, a reasonable representation of the
crystal potential can often be obtained even in open crystals
by adding suitably placed `empty' spheres as basis
`atoms' in the lattice.
Central for our purpose are matrix elements of
the KKR-CPA ensemble-averaged Green function $G(E)$, and the
\hbox{one-site} restricted Green function $G^{A(B)}$
where a specific atom $X_{k_{CP}}$ (A or B) sits
on the central site of the disordered $k_{CP}$-th sublattice,
while all other sites of the $k_{CP}$-th sublattice are occupied by the
effective CPA atom. The relevant expressions are
\cite{ban87,sto84,footnote4,mijnarends76,bansil81}
\begin{mathletters}
\label{eqover:01}
\begin{eqnarray}
 & < s^{'} , {\bf r} ^{'} + {\bf a}_{k_{CP}} \vert G^{A(B)}(E) \vert
     s     , {\bf r}      + {\bf a}_{k_{CP}} >  =
   - \sum\limits_{\sigma L}
       J^{A(B)}_{\sigma L} (s^{'} {\bf r^{'}})
       Z^{A(B)}_{\sigma L} (s     {\bf r    }) +
\nonumber \\
 & + \sum \limits_{\sigma^{'} L^{'}, \sigma L}
       Z^{A(B)}_{\sigma^{'} L^{'}}  (s^{'} {\bf r}^{'} )
       T^{A(B)}_{k_{CP} \sigma^{'} L^{'} , k_{CP} \sigma L}
       Z^{A(B)}_{\sigma     L    }  (s     {\bf r}     ),
\label{eqover:01a}
\end{eqnarray}

\begin{eqnarray}
& < s^{'} , {\bf r} ^{'} + {\bf a}_{k_{CP}} \vert G(E) \vert
     s     , {\bf r}      + {\bf a}_{k_{CP}} >   =
 c_{A}
 < s^{'} , {\bf r} ^{'} + {\bf a}_{k_{CP}} \vert G^{A}(E) \vert
     s     , {\bf r}      + {\bf a}_{k_{CP}} >  +
\nonumber \\
& c_{B}
 < s^{'} , {\bf r} ^{'} + {\bf a}_{k_{CP}} \vert G^{B}(E) \vert
     s     , {\bf r}      + {\bf a}_{k_{CP}} >,
\end{eqnarray}
\label{eqover:01b}
and
\begin{eqnarray}
 & < s^{'} , {\bf r} ^{'} + {\bf a}_{k^{'}} \vert G(E) \vert
     s     , {\bf r}      + {\bf a}_{k    }>  =
  - \sum\limits_{\sigma L}
        J^{(k)}_{\sigma L} (s^{'} {\bf r^{'}})
        Z^{(k)}_{\sigma L} (s     {\bf r    })
        \delta_{kk^{'}} +  \nonumber \\
 &  \sum \limits_{\sigma^{'} L^{'}, \sigma L}
        Z^{(k^{'})}_{\sigma^{'} L^{'}}  (s^{'} {\bf r}^{'} )
        T^{CP}_{k^{'} \sigma^{'} L^{'} , k \sigma L}
        Z^{(k    )}_{\sigma     L    }  (s     {\bf r}     ),
 \; \; {\rm if} \;\; {k \; {\rm and} \; k^{'}\neq k_{CP}} \;.
\label{eqover:01c}
\end{eqnarray}
\end{mathletters}
Here, $r^{'} >  r$,
if $r > r^{'}$ then $J$ and $Z$ should be transposed.
$Z^{(k)}_{\sigma L}$ and $J^{(k)}_{\sigma L}$ are the regular
and irregular solutions, respectively, of the radial Schr{\"o}dinger
equation within the $k$-th muffin-tin sphere, which may be written
compactly as
\begin{eqnarray}
\sum \limits_{s=(+,-)} \{ (E + \nabla^{2}) \delta_{s^{'}s} -
  [v^{(k)}_{0}({\bf r}) \delta_{s^{'}s}  +
 v^{(k)}_{1}({\bf r}) {\bf \hat n} \cdot {\bf \sigma }_{s^{`}s}] \}
  Z^{(k)}_{\sigma L} (s {\bf r})= 0,
\label{eqover:02}
\end{eqnarray}
where the up- and down-spin potentials at the $k$-th site,
$v^{(k)}_{+}$ and $v^{(k)}_{-}$, are combined into a scalar part,
$v^{(k)}_{0}=(v^{(k)}_{+}+v^{(k)}_{-})/2$, and a spin-dependent part,
$v^{(k)}_{1}=(v^{(k)}_{+}-v^{(k)}_{-})/2$. ${\bf \hat n}$ is a unit vector
along the direction of the magnetic moment.
${\bf \sigma}=(\sigma_{x},\sigma_{y},\sigma_{z})$
is a vector composed of Pauli matrices.
$Z^{(k)}$ and $J^{(k)}$ are normalized such that outside
the muffin-sphere (i.e., for $r > S_{k}$ ) they possess the form
\begin{mathletters}
\label{eqover:03}
\begin{eqnarray}
Z^{(k)}_{\sigma L}(s {\bf r})=\sum \limits_{\sigma^{'} L^{'}}
\chi_{\sigma^{'}}(s)j_{l^{'}}(\sqrt{E}r)Y_{L^{'}}(\hat {\bf r})
 [\tau^{(k)}]^{-1}_{\sigma^{'} L^{'},\sigma L} -
i\sqrt{E}\chi_{\sigma}(s)h^{+}_{l}(\sqrt{E}r)Y_{L}(\hat {\bf r}),
\label{eqover:03a}
\end{eqnarray}
\begin{equation}
J^{(k)}_{\sigma L}(s \vec {\bf r})=
\chi_{\sigma}(s)j_{l}(\sqrt{E}r)Y_{L}(\hat {\bf r}).
\label{eqover:03b}
\end{equation}
\end{mathletters}
\noindent
Here $Y_{L}({\bf \hat r})$ is a real spherical harmonic, and $L=(l,m)$
is a composite angular and magnetic quantum number index.
The spin index $\sigma =(+,-)$ and the
spin variable $s=(+,-)$ allow the treatment of lattices
with magnetic ordering.
$\chi_{\sigma}(s)=\delta_{\sigma s}$ denotes the spin part of the
wavefunction. $h^{+}(x)=j_{l}(x)+in_{l}(x)$ is a spherical Hankel
function, where $j_{l}(x)$ is a
spherical Bessel, and $n_{l}(x)$ a spherical Neumann function.
The matrix $\tau^{(k)}(E)$ is built from on-the-energy-shell
elements of the t-matrix of the atom $X_{k}$ on the $k$-th site
(or A, or B atom if $k=k_{CP}$). The elements of
$\tau^{(k)}(E)$ are related to the corresponding
phase shifts $\eta^{(k)}_{\sigma l}(E)$ by
\begin{equation}
\tau^{(k)}_{\sigma^{'} L^{'},\sigma L}(E)=-\sqrt{E}
\exp(i\eta^{(k)}_{\sigma l}) \sin(\eta^{(k)}_{\sigma l})
\delta_{\sigma^{'}\sigma}\delta_{L^{'}L}.
\label{eqover:04}
\end{equation}

The matrix $T^{A}$ (or $T^{B}$) in Eq.~(\ref{eqover:01b})  denotes the
so-called central path operator (in the sublattice-site representation) for
an A or B impurity placed in the KKR-CPA effective medium and
is related to the medium path operator  $T^{CP}$ through the equation
\begin{equation}
T^{A(B)}=T^{CP}[1+(\tau_{A(B)}^{-1}-\tau_{CP}^{-1})T^{CP}]^{-1}.
\label{eqover:05}
\end{equation}
In Eq.~(\ref{eqover:05}), matrix $\tau_{CP}$ is constructed from
atomic matrices  $\tau^{X_{k}}$ on sublattices
\hbox{with $k \neq k_{CP}$}, and from the effective scattering
matrix $\tau^{CP}$ on the $k_{CP}$-sublattice, i.e.,
\begin{eqnarray}
 [\tau_{CP}]_{k^{'}\sigma^{'}L^{'},k\sigma L}=\left\{
\begin{array}{cccl}
\delta_{k^{'}k} & [\tau^{X_{k}}]_{\sigma^{'}L^{'},\sigma L}, &
                                        (k\neq k_{CP}), \\
\delta_{k^{'}k} & [\tau^{CP   }]_{\sigma^{'}L^{'},\sigma L}, &
                                        (k =    k_{CP}).
\end{array}
\right.
\label{eqover:06}
\end{eqnarray}
Similarly, for $\tau_{A(B)}$ we have
\begin{eqnarray}
 [\tau_{A(B)}]_{k^{'}\sigma^{'}L^{'},k\sigma L}=\left\{
\begin{array}{cccl}
\delta_{k^{'}k} & [\tau^{X_{k}}]_{\sigma^{'}L^{'},\sigma L}, &
                                        (k\neq k_{CP}), \\
\delta_{k^{'}k} & [\tau^{A(B) }]_{\sigma^{'}L^{'},\sigma L}, &
                                        (k =    k_{CP}).
\end{array}
\right.
\label{eqover:07}
\end{eqnarray}
The matrix $T^{CP}$ in Eq.(~\ref{eqover:05}) is given by the Brillouin zone
summation
\begin{equation}
T^{CP}_{k^{'}\sigma^{'}L^{'},k\sigma L}=
{1\over{N}} \sum \limits_{{\bf k}\in BZ}
[\tau_{CP}^{-1}-B(E,{\bf k})]^{-1}_{k^{'}\sigma^{'}L^{'},k \sigma L},
\label{eqover:08}
\end{equation}
with
\begin{equation}
[B(E, {\bf k})]_{k^{'}L^{'},k L}=\sum\limits_{{\bf R}_{n^{'}n}}
\exp(i{\bf k}{\bf R}_{n^{'}n})[B(E)]^{(n^{'},n)}_{k^{'}L^{'},k L},
\label{eqover:09}
\end{equation}
which are the KKR-complex crystal structure functions,\cite{segall57}
defined via a multipole expansion of the
free electron Green function $G_0(E)$:
\begin{eqnarray}
 & <{\bf r}^{'}+{\bf a}_{k^{'}}+{\bf R}_{n^{'}} \vert G_{0}(E) \vert
    {\bf r}    +{\bf a}_{k    }+{\bf R}_{n    } >=    \nonumber    \\
 & -i\sqrt{E}\sum\limits_{L}j_{l}(\sqrt{E}r_{<})h^{+}_{l}(\sqrt{E}r_{>})
Y_{L}({\bf\hat r^{'}})Y_{L}({\bf\hat r})\delta_{k^{'}k}\delta_{n^{'}n}+
\sum\limits_{L^{'}L}
Y_{L^{'}}({\bf\hat r^{'}})
[B(E)]^{(n^{'},n)}_{k^{'}L^{'},kL}
Y_{L    }({\bf\hat r    }).
\label{eqover:10}
\end{eqnarray}

The CPA-scattering matrix $\tau_{CP}$ in the Eq.~(\ref{eqover:05})
must be obtained by  solving the KKR-CPA self-consistency
condition
\begin{equation}
c_{A}T^{A}+c_{B}T^{B}=T^{CP}.
\label{eqover:11}
\end{equation}
To solve  Eq.~(\ref{eqover:11}) we use the iteration scheme based on the
following expansion
\begin{eqnarray}
&    [(T^{CP}_{n})^{-1}+
       (\tau_{CP}^{n+1})^{-1}-(\tau_{CP}^{n})^{-1}]^{-1}=
  c_{A}[(T^{CP}_{n})^{-1}+
       (\tau_{A})^{-1}-(\tau_{CP}^{n})^{-1}]^{-1}+ \nonumber \\
&c_{B}[(T^{CP}_{n})^{-1}+
       (\tau_{B})^{-1}-(\tau_{CP}^{n})^{-1}]^{-1}.
\label{eqover:12}
\end{eqnarray}
Equation~(\ref{eqover:12}) allows the computation of $\tau_{CP}^{(n+1)}$
in terms of $\tau_{CP}^{(n)}$ and $T^{CP}_{n}$. By carrying out the
integration of Eq.~(\ref{eqover:08}), $T^{CP}_{n+1}$ is then determined,
and the next iteration cycle can be started. This procedure is rigorously
convergent and preserves the analytic properties of the solutions in
the complex energy plane.\cite{muller73,ducastelle74,mills78,kaplan80}
This is crucially important because many other schemes
used in the literature usually fail at some energy points.
The problem generally becomes more severe as one considers
systems with larger number of atoms per unit cell, and we have found that
Eq.~(\ref{eqover:12})  must be the basis of any robust automated procedure
for obtaining selfconsistent KKR-CPA solutions.

We have now completely defined the computation of the Green function
in Eqs.~(\ref{eqover:01}) for a given crystal potential.
For carrying out charge- and spin-selfconsistency cycles,
one other key parameter, namely, the Fermi energy $E_F$,
must be evaluated.
In this connection, we have developed a powerful version of the
Lloyd formula\cite{tobola92,kaprzyk90,kaprzyk97,zhang92}
for the total number of states below any energy
by formally integrating the trace of the
KKR-CPA Green function exactly in the complex energy plane.
The generalization of this formula to spin-dependent
multi-atom alloys is given below. We emphasize that highly
 accurate charge- and spin-selfconsistent
KKR-CPA results of the sort presented in this article would not be
possible to obtain without the use of this Lloyd-type formula
for determining the alloy Fermi energy. This is an important point
because errors in the Fermi energy determination
at any stage of the computation impede the convergence of
selfconsistency cycles, and degrade the accuracy of the final solution
for charge- and spin-densities as well as other physical properties.

We start by taking the trace of the KKR-CPA Green function
over the spin- and position space, i.e.,
\begin{equation}
G(E)=\sum\limits_{s=(+,-)}
     \sum\limits_{k=1}^{K}
     \int\limits_{V_{k}}d^{3}r
<s,{\bf r}+{\bf a}_{k} \vert G(E) \vert s,{\bf r}+{\bf a}_{k} >.
\label{eqover:13}
\end{equation}
The integral
in Eq.~(\ref{eqover:13}) extends over the Voronoi polyhedron $V_k$
around the $k$-th site, and not the muffin-tin
sphere, so that space is filled up exactly.
Assuming a collinear magnetic structure (same z-axis on each site),
the arguments of Ref.~\onlinecite{kaprzyk90}  can be extended
straightforwardly to prove that
\begin{eqnarray}
& G(E)=
               TrQ +
  \frac{d}{dE} \{ {1\over{N}} \sum\limits_{{\bf k}\in BZ}
               Tr\ln[\tau_{CP}^{-1}-B(E,{\bf k})] \}    - \nonumber \\
&-\frac{d}{dE} \{ \sum\limits_{k\neq k_{CP}}Tr\ln(\phi^{X_{k}})  +
               c_{A}Tr\ln(\phi^{A})+c_{B}Trln(\phi^{B}) \} \nonumber  \\
&+\frac{d}{dE} \{Tr\ln[\tau_{A}^{-1}-\tau_{B}^{-1}]        -
               c_{B}Tr\ln[\tau_{CP}^{-1}-\tau_{A}^{-1}]    -
               c_{A}Tr\ln[\tau_{CP}^{-1}-\tau_{B}^{-1}] \},
\label{eqover:14}
\end{eqnarray}
where $\phi^{(k)}_{\sigma l}(E)$ is an energy-dependent renormalization
factor for wavefunction $Z^{(k)}_{\sigma l}(E,r)$  defined by
\begin{equation}
Z^{(k)}_{\sigma l}(E,r)=
\phi^{(k)}_{\sigma l}(E)\Psi^{(k)}_{\sigma l}(E,r)
\label{eqover:15}
\end{equation}
with $\Psi^{(k)}_{l} \rightarrow r^{l} $ (for $r \rightarrow 0$) .
In Eq.~(\ref{eqover:14}), $TrQ$ is the free electron contribution,
\begin{eqnarray}
TrQ=\sum\limits_{\sigma L } \sum\limits_{k}
   {\int\limits_{V_{k}} d^{3}r
  j_{l}(\sqrt{E}r)Y_{L}({\bf \hat r})
 [-i\sqrt{E}h_{l}^{+}(\sqrt{E}r)Y_{L}({\bf \hat r})]
 +\frac{d}{dE}\ln{(\sqrt{E})^{l}}}.
\label{eqover:16}
\end{eqnarray}

Equation~(\ref{eqover:14}) is cumbersome to use in practical applications as
it involves the on-shell elements of the t-matrices, $\tau$,
which do not extend properly into the complex energy plane; a form
in terms of the logarithmic derivatives turns out to be more useful.
First, we write the logarithmic derivative at the $k$-th muffin-tin sphere as
\begin{equation}
D_{k\sigma l}(E)=S_{k}^{2}
\frac{\partial}{\partial r}\ln{Z^{(k)}_{\sigma l}(E,r)}\vert_{r=S_{k}}.
\label{eq.17}
\end{equation}
The diagonal elements of the $\tau $-matrices
are related to the D-matrices by
\begin{eqnarray}
\tau^{-1}_{k\sigma l}(E)=
\frac{1}{j_{l}(\sqrt{E}S_{k})}
\frac{1}{D_{k\sigma l}(E)-D^{(j)}_{kl}}
\frac{1}{j_{l}(\sqrt{E}S_{k})}
+i\sqrt{E}\frac{h^{+}_{l}(\sqrt{E}S_{k})}{j_{l}(\sqrt{E}S_{k})},
\label{eqover:18}
\end{eqnarray}
where
\begin{eqnarray}
D^{(j)}_{kl}(E)=S^{2}_{k}
\frac{\partial}{\partial r}\ln{j_{l}(\sqrt{E}r)}\vert_{r=S_{k}} .
\label{eqover:19}
\end{eqnarray}
We also require the angular-momentum representation of the
free-electron Green function with position vectors on muffin-tin spheres,
\begin{eqnarray}
& [G_{0}(E, {\bf k})]_{k^{'}\sigma^{'}L^{'},k\sigma L}=
 [D^{(h)}-D^{(j)}]^{-1}
 \delta_{\sigma^{'}\sigma}\delta_{k^{'}k}\delta_{L^{'}L} +
 \nonumber \\
& j_{l^{'}}(\sqrt{E}S_{k^{'}})
 [B(E,{\bf k})]_{k^{'}L^{'},kL}
 j_{l    }(\sqrt{E}S_{k    })
 \delta_{\sigma^{'}\sigma}
\label{eqover:20}
\end{eqnarray}
where
\begin{eqnarray}
 D^{(h)}_{kl}(E)=S^{2}_{k}
\frac{\partial}{\partial r}\ln{h^{+}_{l}(\sqrt{E}r)}\vert_{r=S_{k}} .
\label{eqover:21}
\end{eqnarray}
Eliminating $\tau$ and $B$ in favor of $D$ and $G$
yields our final formula
\begin{eqnarray}
 & G(E)=
   -\frac{d}{dE} \{
    \frac{1}{N}\sum\limits_{{\bf k} \in BZ}
    Tr\ln [G^{-1}_{0}(E,{\bf k})+D^{(j)}-D_{CP}]^{-1} \} \nonumber \\
 & -\frac{d}{dE} \{
    c_{A}Tr\ln[\Psi_{A}^{-1}G^{A}]+c_{B}Tr\ln[\Psi_{B}^{-1}G^{B}]
   -Tr\ln G^{CP} \} \nonumber \\
 & +\frac{d}{dE} \{
    \sum\limits_{k\neq k_{CP}} Tr\ln[\Psi^{(k)}] \},
\label{eqover:22}
\end{eqnarray}
where
\begin{eqnarray}
 G^{A(B)}=[(G^{CP})^{-1}+D_{CP}-D_{A(B)}].
\label{eqover:221}
\end{eqnarray}

Equation~(\ref{eqover:22}) is formally exact and possesses the
form of a perfect derivative.
Although several terms in Eq.~(\ref{eqover:22}) are real on the real
axis, their inclusion is crucially important for obtaining an analytically
correct form which can be used throughout the complex plane.
Equation~(\ref{eqover:22})
not only accounts properly for all physical states, but also
removes contributions
from spurious singularities present in the scattering matrices and
path operators. The formal integration of Eq.~(\ref{eqover:22}),
\begin{equation}
N(E)=-\frac{1}{\pi}Im\int\limits_{-\infty}^{E}dEG(E),
\label{eqover:23}
\end{equation}
immediately gives the total number of states,
including all core states, below $E_F$.
The value of $E_F$ itself corresponds to the condition
$N(E_{F})=Z$, where $Z$ is the total number of electrons in the
Wigner-Seitz cell.

For a collinear magnetic structure, each spin direction can be
treated separately via formula (\ref{eqover:13}) yielding
the spin-resolved DOS function
\begin{equation}
\rho_{\sigma}(E)=\frac{\partial}{\partial E}N_{\sigma}(E).
\label{eqover:24}
\end{equation}
The magnetic moment $\mu$ of a W-S cell can be calculated from
\begin{equation}
\mu=N_{+}(E_{F})-N_{-}(E_{F}).
\label{eqover:25}
\end{equation}
The spin-dependent charge density at the $k$-th site is
\begin{eqnarray}
\rho^{(k)}_{\sigma}({\bf r})=
-\frac{1}{\pi}\int\limits_{-\infty}^{E_{F}}dE
 < \sigma ,{\bf r}+{\bf a}_{k}
   \vert G(E) \vert
   \sigma ,{\bf r}+{\bf a}_{k} >,
\label{eqover:26}
\end{eqnarray}
where, if $k=k_{CP}$, then the $k$-th atom is taken as an A or B
atom and the site-restricted Green function of
Eq.~(\ref{eqover:01a}) is used. The spin density at the $k$-th
site is
\begin{eqnarray}
 s^{(k)}({\bf r})=\rho^{(k)}_{+}({\bf r})-\rho^{(k)}_{-}({\bf r}).
\label{eqover:27}
\end{eqnarray}

The preceding equations allow the computation of
KKR-CPA charge and spin densities in the alloy for a
starting crystal potential. A new crystal potential may then be
constructed via the use of the LSD exchange-correlation scheme.
The iteration of this procedure leads to the fully
charge- and spin-selfconsistent KKR-CPA electronic spectrum.

The magnetic moment on the $k$-th site,
\begin{equation}
\mu^{(k)}=
\mu_{B}\int\limits_{\Omega_{k}}d^{3}r s^{(k)}({\bf r}),
\label{eqover:28}
\end{equation}
is defined as an integral over the muffin-tin sphere volume $\Omega_{k}$.
Note that within the framework of the muffin-tin Hamiltonian,
this is a unique way of defining
site-dependent moments in a multi-component system.
Since the muffin-tin spheres are not
space filling, the sum of such individual
moments will in general not equal the total moment
in the unit cell obtained from Eq.~(\ref{eqover:25}) above,
although the differences in the present case turn out to be
rather small.

%-----------------end section over-------------------------------------

%%%

\section { Structural Aspects }
\label{sec:Stru}
A brief discussion of the salient features of the crystal structure will
help the consideration of the electronic properties in the following
section. The unit cell shown in Fig.~\ref{fig1} is a cube of side $a$ and
may be viewed as consisting of four interpenetrating
fcc lattices denoted by the letters A through D; each atom in fact
sits at the center of a cube of side $a/2$ with corners
occupied by various atoms,
so that the packing is identical to that of a simple bcc lattice.
Fe$_{2}$VX possesses the classic L2$_1$ structure associated usually
with the Heusler compounds. Here, the two Fe atoms occupy equivalent
crystallographic positions A and C, while V sits on B sites, and the metalloid
X on the D sites. The V atoms on the B sites are surrounded by
8 Fe nearest neighbors (nn's) in a bcc arrangement. Each of the
Fe atoms in A or C positions has 4 V nn's (B) and 4 X nn's (D).
The 4 metalloids (D) are located in a relative tetrahedral arrangement with
respect to each other; this is also the case for the 4 V atoms
in the B positions. These remarks make it clear that the Heusler
compounds contain structural units characteristic of metals as
well as semiconductors.

The substitution of V by Fe in Fe$_{2}$VX giving the
disordered alloys Fe$_{3-x}$V$_{x}$X causes no change in the
nearest neighbor environment of the V atoms which continue to
have 8 Fe nn's. The substituted Fe atoms in the B positions,
on the other hand, possess 8 Fe nn's as in bcc Fe,
in sharp contrast to the 4 Fe nn's around each Fe atom in Fe$_{2}$VX;
the end compound Fe$_{3}$X is thus rather close to bcc Fe.
Although Fe$_{3}$X with two chemical species is classified as a DO$_{3}$
structure, note that in a solid state sense Fe$_{3}$X really
contains three different types of `atoms' , i.e., two different Fe `atoms'
and the metalloid. It should furthermore be noted that, although the
alloy Fe$_{3-x}$V$_x$Ga crystallizes in the L2$_1$ structure, the end
compound Fe$_3$Ga has the DO$_3$ structure only in the narrow
temperature range 900~K~$<~T~<$~920~K. Below 900~K the stable phase
for small $x$ is the Cu$_3$Au-type (L1$_2$) structure. The DO$_3$
structure of Fe$_3$Ga is metastable and can be obtained by
quenching.\cite{kawamiya83,kawamiya72}

The fact that V replaces the Fe atoms only
in the B sites of Fe$_{3}$X has been
adduced from NMR, M\"ossbauer and neutron diffraction measurements
(see, e.g., Ref.\onlinecite{niculescu83}).
More generally, in the series Fe$_{3-x}$T$_{x}$Si and
Fe$_{3-x}$T$_{x}$Ga, where T denotes a transition metal,
impurities to the left of Fe in the Periodic Table (Mn, V)
show a strong preference for the B sites while those to the right (Co, Ni)
enter at A or C sites.
Interestingly, Cr seems to distribute almost
randomly at A, B, and C sites in  Fe$_{3-x}$Cr$_{x}$Si.\cite{waliszewski94}
Reference \onlinecite{ishida89} has considered
the question of preferential occupation of various sites
in the Fe$_{3}$Ga matrix
via band-theory based total energy computations.

\section { Computational Details}
\label{sec:Comp}
We have carried out fully charge and spin self-consistent
KKR-CPA computations on the series Fe$_{3-x}$V$_{x}$X,
for $x=$ 0.0, \ 0.1, \ 0.25, \ 0.5, \ 0.75 and 1.0 with the
metalloid X being Si, Ga, or Al.
In the case of the end compounds
Fe$_{3}$X and Fe$_{2}$VX the KKR-CPA results were
verified by extensive computations based on our totally independent
KKR band-structure codes.
The self-consistency cycles were
repeated for each alloy composition until the {\em maximum}
difference between the input and output muffin-tin
potentials was less than 1 mRy at any mesh point in the
unit cell. Therefore the final potentials used in the evaluation of various
physical quantities are highly self-consistent. All calculations
employ a maximum angular momentum cut-off $l_{max}$=2
and the exchange-correlation
functional of the Barth-Hedin form.\cite{barth72}

The four basis atoms
were placed as follows: Fe(A)=(1/4,1/4,1/4), Fe(C)=(3/4,3/4,3/4),
Fe(B) or V(B)=(1/2,1/2,1/2), and X(D)=(0,0,0).
The experimental values of the lattice
constants\cite{kawamiya82,niculescu76,motoya83} of Fe$_3$X
are given in Table \ref{table1}.
The lattice constant decreases
by less than 1$\%$ in going from Fe$_3$Al to Fe$_2$VAl.\cite{popiel89}
In contrast, the lattice constant slightly increases
(less than 0.5$\%$) with increasing V content
in Fe$_{3-x}$V$_{x}$Si. In Fe$_{3-x}$V$_{x}$Ga also
the value of $a$ differs by only about 1$\%$ between
Fe$_{3}$Ga and Fe$_{2}$VGa.\cite{buschow81}
Thus, the composition dependence of the lattice constant in
Fe$_{3-x}$V$_{x}$X is rather weak, and in this work we have
neglected the effect of this variation,
and taken the $a$ value for all compositions
to be the same as that of the end compound Fe$_{3}$X.

The KKR-CPA cycles were carried out in the complex energy plane
using an elliptic contour beginning at the bottom of the valence bands,
and ending at the Fermi energy determined precisely to an accuracy of
better than 0.1 mRy via the generalized Lloyd formula
of Eq.~(\ref{eqover:22}).
This elliptic contour was divided into 12 sections with 4 Gaussian
quadrature points each, and thus contained a total of 48 energy points;
the maximum imaginary part of the energy was 0.25 Ry. The KKR-CPA
Green function in Eqs.~(\ref{eqover:01})
was computed on a 75 special {\bf k} point mesh\cite{froyen89}
in the irreducible part of the Brillouin zone
(BZ) for each of the 48 aforementioned energy points in order to
evaluate the BZ integral of Eq.~(\ref{eqover:08}).
In this way the KKR-CPA self-consistency condition was
solved at each of the 48 basic energy points to an accuracy of about
1 part in 10$^5$, followed by the computation of a new spin-dependent
crystal potential. The starting potential for the next cycle was
typically obtained by a roughly 10$\%$ mixing of the new potential.
The solution of the KKR-CPA condition required 1-10 iterations, while
10-50 charge and spin self-consistency cycles were usually needed depending
upon the alloy composition to achieve     convergence of the
crystal potential to an absolute accuracy of about 1 mRy.
For the final potentials, the total density of states (DOS),
site-decomposed component densities of states (CDOS), and the
{\em l}-decomposed partial densities of states (PDOS) were computed
on a 201 energy point mesh in the alloys, and a 401 point mesh for
the end compounds using a tetrahedral {\bf k} space integration
technique\cite{kaprzyk86}
(with division of 1/48-th of the BZ into 192 small tetrahedra)
applicable to the ordered as well as the disordered muffin-tin systems.

%%%

%--------------------------------------------------------------------------

\section{RESULTS AND DISCUSSION}
\label{sec:Resu}

\subsection{An Overview within a Simplified Model Density of States}
\label{subsec:model}

We present first a relatively simple picture of the component 
densities of states associated with the transition metal atoms in 
Fe$_{3-x}$V$_{x}$X. The model density of states of Fig.~\ref{fig2} gives the
majority (up) and minority (down) spin densities on Fe(A,C), Fe(B), and 
V sites in the limiting cases $x=0$ and $x=1$. The positions of the 
centers of gravity   of the Si $3p$, Fe $3d$ and V $3d$ bands are
shown, together with the Fermi energies ($E_F$) for the 
tetravalent (Si) and trivalent (Ga and Al) metalloids. 
In the following discussion we invoke Fig.~\ref{fig2} frequently
in order to gain insight into the electronic structure and magnetism 
of Fe$_{3-x}$V$_{x}$X. We emphasize however that, even though
the model of Fig.~\ref{fig2} captures a good deal of the physical
essense of the underlying spectrum, the full KKR-CPA
selfconsistent results should always be kept in mind. 
 This is especially true in the 
way Ga and Al-compounds are modeled in Fig.~\ref{fig2} via a rigid
shift in the position of the $E_F$ compared to the case of Si, the 
real situation of course being more complicated. 

The relative positions of various levels in Fig.~\ref{fig2} can be
used to obtain a qualitative handle on the movement of bonding and 
anti-bonding states on various atoms. This aspect will play an 
important role in our analysis below and, therefore, we comment 
briefly on this point with reference to Fig.~\ref{fig3} which
describes the so-called common band model of bonding of $d$-band 
metals.\cite{pettifor95} Consider two atoms, A and B, with 
atomic energy level $E_{A}^0 < E_{B}^0$, which are assumed
to broaden into rectangular bands of common bandwidth $W$ when the atoms
are brought together to form a solid. From moment theory, it is known
that the center of gravity of the local density of states (occupied
and unoccupied) must coincide with the local on-site
energy level $E_{A(B)}$ (which may be slightly
shifted from the corresponding free-atom value in order to maintain
local charge neutrality). The result is a skewing of the
originally rectangular local density of states and a
new bandwidth $W_{AB}$. Physically, the skewing represents
a transfer of charge from the B atom to the A atom until the Fermi
levels become the same. With this redistribution,
the states at the bottom of the band (the bonding
states) become more concentrated on the A atom while the antibonding
states at the top of the band are found preferentially on the B atom
as shown in Fig.~\ref{fig3}b. This mechanism constitutes a
basic  ingredient for understanding the electronic structure and
magnetism of Fe$_{3-x}$V$_{x}$X considered below.
A similar discussion based on the bonding of molecular orbitals
has been given by Williams {\em et al.}\cite{williams82} 

We are now in a position to consider the behavior of the total 
magnetic moment and its constituent parts in Fe$_{3-x}$V$_{x}$X 
(Figs.~\ref{fig4}-\ref{fig6}) in terms of the simple model of
Fig.~\ref{fig2}. Some salient
features which may be explained are as follows.

(i) {\em Smaller moment of Fe(A,C) compared to Fe(B) in Fe$_3$X}.
Since the Fe(B) atom is surrounded by eight Fe(A,C) atoms in a bcc
arrangement, it is not surprising to find the component density
of states (CDOS) for the $d$ electrons of Fe(B) in Fig.~\ref{fig2}a to
show the familiar structure of two peaks (bonding and
anti-bonding) separated by a valley of low density of states
found in bcc metallic Fe. For the up-spin electrons both peaks are 
occupied, while the exchange splitting pushes the down-spin antibonding 
peak above the Fermi level, resulting in a large magnetic moment on 
the Fe(B) site. 

Concerning the Fe(A,C) moment, note first that 
Fe(A,C) is coordinated with four Fe(B) and four metalloids,
and that the associated CDOS in Fig.~\ref{fig2} possesses
extra states between the bonding and antibonding peaks. The main difference 
in relation to Fe(B) is in the behavior of the {\it down} spins for which 
Fig.~\ref{fig2}a shows that $E_{\mbox{\scriptsize Fe(A,C)}} <
E_{\mbox{\scriptsize Fe(B)}}$. Therefore, down-spin bonding states 
reside preferentially on Fe(A,C) and yield an increased negative 
spin density compared to Fe(B); the corresponding antibonding states
on Fe(B) lie above $E_F$ and are therefore unoccupied.
Fe(A,C) and Fe(B) are quite similar 
with respect to the up spins since bonding as well as anti-bonding 
states on both Fe sites lie below the Fermi energy. 
The net result is that the total moment
on Fe(A,C) is reduced compared to Fe(B). 

(ii) {\em The negative moment of a V-impurity in Fe$_3$X}. We see from 
Fig.~\ref{fig2}a that $E_{\mbox{\scriptsize V}}$ is
higher than $E_{\mbox{\scriptsize Fe(A,C)}}$
and its next-nearest neighbor $E_{\mbox{\scriptsize Fe(B)}}$ for the
up-spin electrons, and thus the up-spin bonding states will move 
away from V sites; the antibonding states will be on V(B), but 
these lie mostly above the Fermi level and are thus unoccupied. 
On the other hand, for down spins $E_{\mbox{\scriptsize V}}$ 
lies somewhat below $E_{\mbox{\scriptsize Fe(A,C)}}$
and $E_{\mbox{\scriptsize Fe(B)}}$. Both effects will tend 
to induce a negative moment on V impurities. 

(iii) {\em The negative moment of Si in Fe$_3$X}. 
The Si atoms too carry a negative moment, albeit small. The
reason is that for the down-spin electrons $E_{\mbox{\scriptsize Si}}$ 
lies well below $E_{\mbox{\scriptsize Fe(B)}}$, 
while for the up-spin electrons $E_{\mbox{\scriptsize Si}} >
E_{\mbox{\scriptsize Fe(B)}}$. As a result of $p-d$ hybridization 
the down-spin bonding states will
be more predominant on Si while the (empty) antibonding states
will be found on Fe(B), resulting in a negative Si moment.

(iv) {\em Positive polarization of Fe(B) impurity in $Fe_2$VX}.
Similar arguments explain why an Fe(B) impurity in Fe$_2$VSi, at
the other end of the concentration range, is positively
polarized. Figure~\ref{fig2}b shows that $E_{\mbox{\scriptsize Fe(B)}} <
E_{\mbox{\scriptsize Fe(A,C)}}$ for the up-spin electrons and thus the bonding
states will be found predominantly on Fe(B). The
opposite situation is true for the down-spin electrons.
Both effects conspire to produce a strong positive Fe(B) moment.

(v) {\em Higher moment of Fe(A,C) but not Fe(B) 
in the Ga- and Al-compound compared to the Si-compound}.
Going to Fe$_3$Ga(Al) we find in Figs.~\ref{fig5} and \ref{fig6}
that, compared to Fe$_3$Si, the Fe(A,C) moments are higher. Figure~\ref{fig2}a
shows why. Ga and Al are trivalent and therefore the Fermi energy
is lower. This does not affect the spin-up band which lies in its
entirety below $E_F$, nor does it affect the Fe(B) spin-down band
because there are hardly any states on Fe(B) in the region
between the bonding and antibonding peaks. The net effect is a
reduction of the number of spin-down electrons on Fe(A,C) with a
resulting larger moment on Fe(A,C), and a negligible effect on the Fe(B)
moment. 

The preceding discussion is meant to be 
illustrative rather than exhaustive. Other features of 
the behavior of moments in Figs.~\ref{fig4}-\ref{fig6} can be understood
at least qualitatively along these lines.

%%%

\subsection{Ordered compound Fe$_3$Si}
\label{subsec:ResFe3Si}

After the introductory discussion of the electronic band
structure of the Fe$_{3-x}$V$_x$X system on the basis of the
simple model introduced in the previous section we now turn to
the calculated density-of-states curves in Fig.~\ref{fig7}. A
good understanding of Fe$_3$Si is essential for delineating the
effects of substitution on the metalloid and/or the Fe(B) site
considered in the following sections. Note that the Si $3s$ and $3p$ 
bands show little overlap (Figs.~\ref{fig7}c4 and \ref{fig7}c3)
in Fe$_3$Si, even though in Si the $3s$ and $3p$ bands possess a substantial 
overlap; this is the result of an increased Si-Si distance in Fe$_3$Si 
compared to Si. 

The bonding between Si and Fe is complex and involves $s$ and $p$
electrons of Si and $s, p$ as well as $d$ electrons of Fe. 
Some manifestations of this bonding are as follows. 
The Si $3s$ states form a semicore band extending below $\sim
0.2$ Ry; the presence of a finite density of states on both 
Fe sites in this energy region indicates Si-Fe interaction involving 
Si $3s$ electrons, even though there are no Si atoms in the Fe(B) 
nn shell. Si-Fe binding via Si and Fe $p$ electrons is 
apparent from the presence of the three-peak structure in the $p$
bands, which is most clearly discernable in the Si down-spin $p$ band
(Fig.~\ref{fig7}c3), but is also present in the up- and down-spin
$p$ bands of both types of Fe sites. Finally, there is the effect of
hybrid formation between the $3p$ states of Si and the $d$ states of
Fe described by Ho {\em et al.}\cite{ho80} and discussed in 
Ref. \onlinecite{williams82} in connection with Fe$_3$Si, 
which tends to concentrate
$p-d$ bonding states on Fe(B) and enhance the moment on Fe(B).

The behavior of $N(E_F)$, the density of states at $E_F$, deserves comment.
The states at $E_F$ possess mostly $d$ character with some $p$ admixture
(Fig.~\ref{fig7}). The Fe(A,C) contribution dominates, with the
spin-down part being much larger than the spin-up part (Fig.~\ref{fig7}a1).
By contrast, the Fermi level on Fe(B) lies in a fairly low density of 
states region in the up- as well as the down-spin CDOS. The energy dependence
of the CDOS in the vicinity of $E_F$ is also quite different on various 
sites. On Fe(A,C), the Fermi level lies near a dip in the up-spin CDOS, 
but in a rapidly decreasing region in the down-spin CDOS (Fig.~\ref{fig7}a1);
a rigid upward shift of 0.04 Ry in $E_F$, for example, would cause a 
reversal of spin polarization at $E_F$. The situation for Fe(B), on 
the other hand, is quite the opposite in that a similar shift in $E_F$ 
will induce a rapid increase in the down-spin density (Fig.~\ref{fig7}b1).

Our computed $\ell$-decomposed Si-CDOS (Fig.~\ref{fig7}c) gives insight
into the Si L$_{2,3}$ soft x-ray emission (SXE) spectrum of Fe$_{3}$Si
reported by Jia {\em et al.}\cite{jia92} The SXE data from Fe$_{3}$Si 
(see Fig.~1 of Jia {\em et al.}) display three distinct peaks centered at 
binding energies of 2 eV, 6 eV, and 10 eV. To interpret these results 
recall first that the L$_{2,3}$ SXE will only involve $s$ and $d$ but 
not the $p$ partial density due to the $\Delta\ell=\pm 1$ selection 
rule for optical transitions. 
The $d$ PDOS (Fig.~\ref{fig7}c2) contains many features extending from 0-6 eV
below $E_F$. In Fig.~\ref{fig8} we have plotted the sum of $s$ and $d$
PDOS for Si after smoothing the theoretical spectrum to reflect experimental 
broadening.\cite{footnote6} The three peaks in Fig.~\ref{fig8} at binding
energies of
1.7 eV, 5.5 eV, and 9.2 eV, are seen to be remarkably consistent with 
the experimental values quoted above. In particular, our calculations 
suggest that the 6 eV peak in the SXE spectrum involves Fe-Si bonding 
$d$ states since both Fe(B) and Fe(A,C) CDOS's possess a substantial 
density in this energy region (Figs.~\ref{fig7}a1 and \ref{fig7}b1,
note the scale); in contrast, Ref.~\onlinecite{jia92} associates 
this peak with $sp^3$ bonded Si orbitals.\cite{footnote1}

%%%

\subsection{Ordered compounds Fe$_3$Ga and Fe$_3$Al}
\label{subsec:ResFe3GaFe3Al}

Fe$_{3}$Si and Fe$_{3}$Ga are compared first (Fig.~\ref{fig9}).
The replacement of Si by Ga is seen to induce only small changes in the 
shape of the down-spin CDOS on either Fe(A,C) or Fe(B), aside from
a relative lowering of the Fermi level to account for the reduced 
valence of the metalloid. The effect on the  up-spin CDOS, on the
other hand, is more substantial in that the shape of both Fe(A,C) and
Fe(B) is flatter around $E_F$ in Fe$_{3}$Ga compared to Fe$_{3}$Si
(e.g., Figs.~\ref{fig9}a2 and \ref{fig9}b2); the Fe-Ga interaction
pulls the up-spin states peaking around the Fermi level in Fe$_{3}$Si
to lower energy.

The spectra for Fe$_{3}$Ga and Fe$_{3}$Al in Fig.~\ref{fig9}
are quite similar. As already noted, the Si $s$ band in Fe$_{3}$Si
(below $\sim$ 0.2 Ry) is more or less core-like; the Ga $s$ band
lies at a lower binding energy closer to the bottom of the valence 
band in Fe$_{3}$Ga. This progression continues in Fe$_{3}$Al where 
the Al $s$ band overlaps the bottom of the valence band causing a 
relatively greater distortion of states in this energy region
(see bottom row in Fig.~\ref{fig9}).

It is noteworthy that substitution
on the Si site by Ga and Al not only influences the Fe(A,C)
but also the Fe(B) CDOS. This is consistent with our observation
above in connection with Fig.~\ref{fig7} that the metalloid affects
the Fe(B) CDOS even though there are no metalloid atoms in the
Fe(B) nn-shell. It is clear, therefore, that a simple `environmental'
type model\cite{niculescu83,switendick76} which is based on taking 
account of {\em only} the composition of the nn-shell possesses intrinsic 
limitations in describing the electronic structure and magnetism of 
Fe$_{3}$X.

%%%

\subsection{Ordered compounds Fe$_2$VX (X=Si, Ga, Al)}
\label{subsec:ResFe2VX}

The substitution of V for Fe(B) in Fe$_3$Si is seen by comparing
Figs.~\ref{fig9}a and \ref{fig10}a to dramatically alter the density of
states. The substitution of the high-moment Fe atom by nonmagnetic V
causes the magnetic moment to nearly collapse. This is reflected in
somewhat different ways in the up- and down-spin densities. Similar to
Fe(B), the V atom in a bcc environment displays the familiar $d$ band
structure of two peaks separated by a region of low density, but since
V is nonmagnetic the exchange splitting is close to zero
(Fig.~\ref{fig10}a3).
In order to accommodate the reduced valence of V compared to
Fe, both the up- and down-spin antibonding V $d$ states are
pushed above $E_F$. The net changes
in relation to Fe$_3$Si are
most clearly visible for the up-spin electrons. The up- and down-spin
densities of Fe$_2$VSi possess similar shapes with the small magnetic
splitting localized essentially on Fe(A,C).
                                            The dip in the up-spin
Fe$_3$Si DOS around 0.6 Ry in Fig.~\ref{fig9}a1, now moved up on the
energy scale, is partially filled by nnn Fe states around 0.75 Ry
(Figs.~\ref{fig10}a1 and \ref{fig10}a2). To the left of these there is
a complex of Fe-V bonding states followed by metalloid $p-d$ bonding
states (Figs.~\ref{fig10}a2 and \ref{fig10}a3).
                                    Between the Fe states and the
antibonding Fe-V states around the Fermi level a near gap has formed
for both spin directions.

The character of states at the Fermi level differs greatly between
Fe$_2$VSi and Fe$_3$Si; the B site (i.e., V(B) {\em vs} Fe(B))
                                                contribution is larger
in the former compared to the latter. The DOS at $E_F$ is much larger in
Fe$_2$VSi and is dominated by Fe(A,C) and V(B) up spins, while
in Fe$_{3}$Si the Fe(A,C) down spins dominate with other contributions
being small. Other differences are evident from
Figs.~\ref{fig9}a and \ref{fig10}a; for example, $E_F$ lies
in up-spin Fe(A,C) and V(B) peaks in Fe$_2$VSi, but in an up-spin
Fe(A,C) dip in Fe$_{3}$Si. Therefore, we should expect Fe$_2$VSi
to respond very differently to rigid shifts of the Fermi energy.

Turning to Fe$_2$VGa (Fig.~\ref{fig10}b), the spectra are quite similar
in shape to Fe$_2$VSi, although the minimum in the DOS around 0.8 Ry
is somewhat deeper and broader in the Si compound. That the spectrum
of Fe$_2$VGa does not possess an actual band gap but only a pseudogap
for either spin direction is seen more clearly from Fig.~\ref{fig11}.
The small downward shift of the Fermi level due to the lower valence of Ga
places the Fermi level firmly in the (pseudo)gap region, thereby precluding
moment formation. In sharp contrast to the Si compound, therefore, the
DOS at $E_F$ in the Ga case is nearly zero for up- as well as
down-spins and would increase rapidly by a rigid lowering or raising
of the Fermi level.

Notably, the gap between the semicore band around 0.15 Ry and the bottom
of the valence band in Fe$_2$VSi is larger than in Fe$_{3}$Si.
This effect is present also in Fe$_2$VGa, and is a consequence of
changes in the various interactions and not due to a change in the
lattice size since we have used a lattice constant independent
of V concentration in our computations. The results for Fe$_2$VAl are seen
to be similar to those for Fe$_2$VGa.
However, in contrast to the case of Fe$_3$Al
(Fig.~\ref{fig9}c), the bottom of the valence band does not overlap the
semicore band around 0.25 Ry in Fe$_2$VAl (Fig.~\ref{fig10}c).

\subsection{Disordered alloys Fe$_{3-x}$V$_x$X (X=Si, Ga, Al)}
\label{subsec:ResDis}
\subsubsection{Fe$_{3-x}$V$_x$Si}

Fe$_{3-x}$V$_x$Si is considered first with the help of
Fig.~\ref{fig12}. The basic effects outlined in the preceding
subsection in connection with Fe$_2$VSi are of course at play here.
Since the driving mechanism is the replacement of Fe(B) by V(B), the b
and c panels in Fig.~\ref{fig12} will be considered first. At $x=0$
(Fig.~\ref{fig12}b1) the Fe(B) site displays the two-peaked structure
discussed earlier. A V impurity on that site however shows already the
upward shift of the two peaks in the up-spin $d$ band. This shift,
together with the interaction of the V atom with its Fe(A,C) neighbors,
has been shown in Subsection \ref{subsec:model} above to lead to the negative
moment on the V impurity.

As the V concentration increases, the Fe(B) density of states shows
the familiar blurring caused by disorder scattering in the alloy. This
effect is however highly non-uniform in that some states are broadened
much more than other states, which reflects large variations in the
effective disorder parameter as a function of {\bf k}, $E$, and spin
polarization.\cite{footnote7} Also, with increasing $x$ the down-spin
bonding peak at 0.7 Ry decreases in size (Fig.~\ref{fig12}b),
leaving an even larger uncompensated moment in the up-spin band. In
the dilute Fe impurity limit (Fig.~\ref{fig12}b5) the moment on this
Fe(B) atom is found to be 3.08 $\mu_B$. It is interesting to see that
in this limit the Fe(B) up-spin bands are virtually undamped again;
they have the character of relatively sharp impurity levels.

As a result of the interaction with the V and Fe atoms on the B sites,
the Fe(A,C) CDOS undergoes quite substantial changes.
With increasing V content, the highest occupied peak around
0.8 Ry in the up-spin Fe(A,C) CDOS moves to higher energies, which
helps deepen the low density of states region near the Fermi level and
pushes $E_F$ to higher values. These large movements in the spectral
weights are not present in the down-spin Fe(A,C) CDOS although the
development of the pseudogap takes place in this case also. The net
result of the spectral weight shifts is a decrease of the up-spin
moment and a simultaneous increase of the down-spin moment due to the
upward shift of the Fermi level. The two effects conspire to cause a
rapid depolarization of the Fe(A,C) sites. For $x \ge 0.5$ the
aforementioned effects more or less saturate and there is little
further change in the various moments.

The magnetic moments associated with different sites are seen from
Fig.~\ref{fig4} to deviate from straight lines joining the $x=0.0$ and
$x=1.0$ values. The moments change essentially linearly for $x \le
0.5$ with the Fe(B) component increasing while the Fe(A,C) component
and the absolute value of the negative moment on V(B) decrease. For
larger V concentrations, all contributions are nearly flat.
Essentially, as Figs.~\ref{fig12} and ~\ref{fig2} show, there
are three interfering mechanisms in going from Fe$_3$Si to
Fe$_2$VSi: (i) Replacement of high-moment Fe(B) by
low-(negative)-moment V(B) atoms; (ii) a gradual upward movement
of the up-spin Fe(A,C) antibonding orbitals which reduces the
Fe(A,C) moment, and (iii) in going from Fe$_3$Si to Fe$_2$VSi, 
as the number of Fe(B) atoms decreases, the Fe(A,C) atoms play an
increasingly important role in bonding. The
concentration dependence of the magnetic moments and their lack
of linearity results from an interplay of these factors.

We have made an extensive comparison of our results on
Fe$_{3-x}$V$_{x}$Si with those of Kudrnovsky {\em et al.}\cite{kudrnovsky91}
Despite some overall similarities, our results differ substantially
from those of Ref. \onlinecite{kudrnovsky91}.
For example, the Fermi level in our case lies
at or near a dip in the {\em majority}  spin-DOS for up to 50$\%$ V
(Fig.~\ref{fig12}a), while in Ref. \onlinecite{kudrnovsky91}
(see their Fig. 8),
the majority spin $N_{+}(E_F)$ decreases
with increasing energy in the 0$\%$
and 25$\%$ alloy and is essentially at a minimum in the 50$\%$ V case.
The Fermi level in Ref. \onlinecite{kudrnovsky91} rises uniformly
 with increasing V content.
In sharp contrast, our $E_F$ changes non-linearly, with the
$E_F$ values being rather close for 0$\%$ and 25$\%$ V
(Figs.~\ref{fig12}a1 and \ref{fig12}a2).
The site-dependent CDOS's show differences as well.
The shape of the Fe(B) majority spin-CDOS is seen from
Fig.~\ref{fig12}b to change considerably in the energy region lying
a few eV's below $E_F$. In Ref. \onlinecite{kudrnovsky91}
(see their Fig. 10), on the other hand, the Fe(B) majority-spin CDOS possesses
a roughly composition independent three peaked shape.
The Fe(B) CDOS at $E_F$ in our computations is quite small
for all V concentrations, while in Ref. \onlinecite{kudrnovsky91}
the minority-spin Fe(B) CDOS at $E_F$ is quite large (see their Fig. 9).
Turning to magnetic moments, we find (Fig.\ref{fig4}) the Fe(B) moment
to {\em increase} with V content, whereas
Ref. \onlinecite{kudrnovsky91} obtains
a {\em decreasing} Fe(B) moment.
Further, our moments on all sites vary linearly up to 50$\%$ V and remain
virtually constant thereafter, and the total moment shows
a related break in slope around 50$\%$ V. The results of
Ref. \onlinecite{kudrnovsky91} do not display these
effects clearly as all moments
appear to vary roughly linearly up to 75$\%$ V.
The comparisons of this paragraph make it clear that the
charge selfconsistency achieved in the present work has
important consequences for the electronic spectrum;
as already noted, the results of Ref. \onlinecite{kudrnovsky91}
are based on a non-selfconsistent crystal potential.

\subsubsection{Fe$_{3-x}$V$_{x}$Ga and Fe$_{3-x}$V$_{x}$Al}

The evolution of the electronic spectrum of Fe$_{3-x}$V$_{x}$Ga
(Figs.~\ref{fig13})
can be understood along much the same lines as Fe$_{3-x}$V$_{x}$Si,
keeping in mind of course the differences in the spectra of the
end compounds discussed already in Sections \ref{subsec:ResFe3Si}
and \ref{subsec:ResFe3GaFe3Al}  above.
Interestingly, there are differences between the composition
dependence of the magnetic
moments between the two compounds as seen by
comparing Figs.~\ref{fig4} and \ref{fig5}.
In Fe$_{3-x}$V$_{x}$Si, the computed moments on all sites remain
essentially constant for $x\ge$0.5, whereas in Fe$_{3-x}$V$_{x}$Ga
the moments on Fe(A,C) and V(B) continue to decrease, and that on Fe(B)
continues to increase with $x$. These effects can be related to the
behavior of the underlying spectra as follows.
Concerning Fe(A,C), note that changes in the Fe(A,C) CDOS for either
spin direction are quite similar in the Ga and
the Si compound insofar as the development
of the pseudogap around the Fermi level and the shifts of spectral
weights in the majority spin CDOS
(Figs.~\ref{fig12}a and \ref{fig14}a) are concerned.
In Fe$_{3-x}$V$_{x}$Si, the Fermi level lies in the down-spin pseudogap
for $x\ge$0.5. But, the smaller valence of Ga (compared to Si) causes
$E_F$ in Fe$_{3-x}$V$_{x}$Ga to be relatively lower.
The minority spin contribution to the Fe(A,C) moment then
continues to increase for $x\ge$0.5 in Fe$_{3-x}$V$_{x}$Ga,
with a concomitant decrease in the total Fe(A,C) moment.

The composition dependence of the electronic spectrum as well as
the magnetic moments on various sites in Fe$_{3-x}$V$_{x}$Al is
quite similar to that of Fe$_{3-x}$V$_{x}$Ga, minor differences
notwithstanding. While the moments in Fe$_{3-x}$V$_{x}$Al are
presented in Fig.~\ref{fig6}, the detailed results for Fe(A,C), Fe(B),
and V(B) CDOS's are not shown in the interest of brevity.

\subsection{Other aspects}
\label{subsec:OthAs}

\subsubsection{Rigid band model {\em vs} common-band model}

Although some features of the Ga and Al alloys can be understood
reasonably via a rigid band picture (see discussion of Fig.~\ref{fig2}
above), our results indicate that
a description of the electronic spectrum of Fe$_{3-x}$V$_{x}$X in
terms of any simple rigid-band-type model is generally unjustified. 
The end compounds Fe$_3$X and Fe$_2$VX possess quite
different spectra for any given metalloid X, and the shapes of the
CDOS's at various sites change with Fe/V substitution. Furthermore,
the fact that the up- and down-spin DOS's possess very different
shapes, particularly in the Fe$_3$X limit, implies that the
ferromagnetic state cannot be described properly to be the result of a
simple spin splitting in the form of a Stoner shift of more or less
rigid paramagnetic bands. However, from our discussion
it is clear that the common-band model constitutes a fertile
framework for obtaining insight in the electronic structure
and magnetism of these materials.
 While the exchange splitting on the Fe(B) atom constitutes
the driving force for magnetism in these compounds, the hybridization
between different states and the resulting differences between the
weights of various bonding and antibonding
states on different atoms lead to a rich variety of
behaviors.

\subsubsection{Site selectivity}

As pointed out in Section~\ref{sec:Stru}, the issue of site
selectivity in the Heusler-type alloys has been the subject of
numerous studies,
\cite{niculescu83,switendick76,garba86,ishida89,budnick89} with the
attention being focused mostly on determining whether other metal
atoms when substituted for Fe preferentially occupy Fe(B) or Fe(A,C)
sites in the lattice. The high electronegativity of Si implies that Si
attracts electrons from the surrounding Fe atoms. Our calculations
indicate that in Fe$_3$Si the number of electrons inside the
muffin-tin spheres of Fe(A,C) is 25.115 against 24.975 for Fe(B),
i.e., Fe(A,C) {\em averaged over both spin directions}
              is more electronegative than Fe(B). The same is true in
Fe$_3$Ga and Fe$_3$Al, even though Ga and Al are less electronegative
than both Fe(A,C) and Fe(B). The Coulomb energy of the crystal lattice
will be reduced if electronegative Fe(A,C) is replaced by an element
more electronegative than Fe (i.e., Co or Ni) or if Fe(B) is replaced by a less
electronegative element (i.e., Ti, V, Cr, or Mn), thus explaining the
observed site selectivity. A more quantitative demonstration of these
effects, of course, must be based on total energy calculations where
the LDA has proved a sound basis for metals,\cite{mjw78}
compounds,\cite{williams82,kubler83} and binary alloys.\cite{johnson86}
Accordingly, we have computed the total energy of the alloy
Fe$_{3-x}$V$_x$Si for a
number of concentrations $x$ putting V atoms first on the
A or C sublattice and subsequently on the B sublattice.
Figure~\ref{fig15} shows the difference between the total energy
for each of these two situations and the sum of the energies of the
constituent atoms for $x \leq 0.08$. It is clear
that the total energy is lowered when the V atom occupies a B site.

\subsubsection{Electrical resistivity}

Figures~\ref{fig12}-\ref{fig14} imply that the type, spin,
and number of carriers available for transport in
the Fe$_{3-x}$V$_{x}$X alloys depend
strongly on the V content as well as on the metalloid valence.
Focusing on Fe$_{3-x}$V$_{x}$Si first,
we see from Fig.~\ref{fig12} that in Fe$_3$Si (topmost row) the
density of states at the Fermi level is dominated by the down-spin
Fe(A,C)-CDOS. Therefore, the current in Fe$_3$Si
will be carried primarily by down-spin electrons associated with
Fe(A,C) sites with relatively little contribution from Fe(B) and
V(B) sites. With increasing V content, the Fermi level in
Fe$_{3-x}$V$_{x}$Si moves into the pseudogap in the down-spin
Fe(A,C)-CDOS; for $x$=0.50, these electrons
are seen to be essentially frozen out of the transport processes.
In fact, for $x$=0.50 few carriers of
either spin are available and one expects the material
to possess a high resistivity. For $x$=0.75, the up-spin Fe(A,C)-CDOS
is larger, and the up-spin V-CDOS begins to increase. In the limiting
case of Fe$_2$VSi (Fig.~\ref{fig12}, bottom row), we see that the current
will be carried mainly by up-spin Fe(A,C) and V(B) electrons. Thus, in
going from Fe$_3$Si to Fe$_2$VSi, the carriers change from being
dominated by down-spin Fe(A,C) electrons to up-spin Fe(A,C) and V(B)
electrons, and the material goes through a high-resistivity range for
intermediate compositions. Experiment confirms this picture:
measurements of the residual resistivity as a function of $x$ in
Fe$_{3-x}$V$_{x}$Si by Nishino {\em et al.}\cite{nishino93} show a
pronounced maximum at $x \sim 0.4$.

The situation with Fe$_{3-x}$V$_{x}$Ga is seen from Fig.~\ref{fig14}
to be similar to that of Fe$_{3-x}$V$_{x}$Si, except that the compound
does not display an intermediate range with few carriers around
$x$=0.50. Instead, the carriers continue to be dominated by down-spin
Fe(A,C) electrons for all compositions, with the total number of
available carriers going nearly to zero in the $x$=1.0 limit. There
seems to be a curious effect in the $x=0.75$ case (Fig.~\ref{fig14}b4)
in that the Fe(B) CDOS displays a substantial up-spin component,
suggesting that this alloy may show unusual transport phenomena.
 
Returning to the compounds Fe$_2$VAl
and Fe$_2$VGa, we note that the Fermi surface
of Fe$_2$VGa is very similar to that of Fe$_2$VAl.\cite{singh98,guo98}
It consists of three small hole pockets at $\Gamma$ and an electron
pocket at X (see Fig.~\ref{fig11}), resulting in a very small number of
carriers.~\cite{footnote5} Recent calculations for Fe$_2$VAl which
include spin-orbit coupling\cite{weht98} find a further reduction in
carrier density. This would make these compounds semimetals or
even semiconductors, although the latter would disagree with a 
Fermi cut-off reported by Nishino {\em et al.}\cite{nishino97} in
Fe$_2$VAl using high-resolution photoemission. Another cause for the
high and strongly temperature and composition dependent resistivity
could be strong electron scattering by spin fluctuations. Although we
find Fe$_2$VAl and Fe$_2$VGa to be virtually
non-magnetic in the V-rich regime, we have
seen that they may nevertheless contain sizable, relatively isolated,
local moments in the form of Fe atoms on V(B) positions, accompanied
by non-stoichiometry or frozen-in antisite defects. The band structure
of these compounds consists of a set of rather flat Fe and V $d$ bands
closely below and above the Fermi level.
A similar situation involving $f$ bands is known to lead to
heavy-fermion behavior in a variety of intermetallic compounds. Also,
the temperature dependence of the resistivity in Fe$_2$VAl is
qualitatively very similar to that of well-known heavy-fermion
compounds such as CeCu$_2$Si$_2$ and CeCu$_6$.\cite{fisk88} Thus, the
resistivity behavior in Fe$_2$VAl and Fe$_2$VGa may be a signature of
heavy-fermion behavior in a $3d$ intermetallic compound. It is also
interesting to note that the concentration $x=0.6$ at which
Fe$_{3-x}$V$_x$Si shows similar resistivity behavior\cite{nishino93}
lies in the concentration range where the Fermi level is at or very
near the strong dip in the density of states which for $x \rightarrow
1.0$ develops into the pseudogap. This suggests that a low carrier
density, possibly in combination with strong local moments, is a
necessary ingredient for semiconductorlike resistivity behavior in
these alloys and compounds. In this respect resistivity measurements
in Fe$_{3-x}$V$_x$Si for $x > 0.6$ would be of interest.

\subsubsection{ Magnetic moments--comparison with experiment}

We comment now on the magnetic moment data given in Table \ref{table2}
for the end compounds.
\cite{moss72,singh98,guo98,garba86,ishida89,ishida76,williams82,kudrnovsky91,kawamiya82,pickart61,paoletti64,hines76}
The experimental data in this table is not meant to be
exhaustive; we focused on sources
which report site-dependent magnetic moment information. Fe$_3$Si has been
investigated most extensively. The situation with the Fe(B) moment
in Fe$_3$Si appears to be good in that the experimental
values lie around 2.2-2.4 $\mu_B$, and are in reasonable accord with
the theoretical value of about 2.5$\mu_B$; the
value of 1.9$\mu_B$ is based on a tight-binding
model computation. A similar level of agreement is also seen generally
on the Fe(B) site in Fe$_3$Ga and Fe$_3$Al.
In the case of the Fe(A,C) moment, on the other hand, the situation
is somewhat less satisfactory as the experimental as well as
theoretical values are more scattered. Theoretically, the metalloid is
predicted to exhibit a weak negative polarization, but
no experimental results are available. Lack of experimental
data is also evident for the magnetic moments in Fe$_2$VX;
notably, experiments of Ref.\onlinecite{kawakami93} give a near zero Fe(A,C)
moment in Fe$_2$VSi in disagreement with computed values
of 0.43 $\mu_B$ (this work),
and 0.37 $\mu_B$ by Ref.\onlinecite{kudrnovsky91}. Interestingly,
Endo {\em et al.}\cite{endo95} report antiferromagnetic
ordering in Fe$_2$VSi; on this basis,
an Fe(A,C) moment of
0.5-1.0 $\mu_B$ has been deduced.\cite{kawakami95}
Of course, it should be remembered that our calculations do not include
an antiferromagnetic ground state.

Also for the alloys Fe$_{3-x}$V$_{x}$X 
we are not aware of any experimental work addressing
contributions of individual sites to the total moment.
In Fe$_{3-x}$V$_{x}$Si (Fig.~\ref{fig4}), the computed total
moments are rather close to the measurements of
Ref.\onlinecite{niculescu83},
although further experimental work in the V-rich alloys, where the
total moment curve is predicted theoretically to display a significant
change in slope, would be interesting. The experimental data in the
Ga\cite{kawamiya72} and Al\cite{tuszynski90} compounds
(Figs.~\ref{fig5} and \ref{fig6}) extend over
a wider composition range and here again there is a reasonable level
of accord with the theory, even though the experimental points are
generally lower, especially for the 50$\%$ V alloy. We should keep
in mind however that our calculations assume a perfectly random
occupation of Fe(B) sites by V atoms while the physical system may
display short range ordering and clustering effects in addition to
possible uncertainties associated with site occupancies.

%%%
%--------------------------------------------------------------------------
\section{Summary and Conclusions}

We have discussed the electronic structure of Fe$_{3-x}$V$_{x}$X Heusler
alloys, where the metalloid X is Si, Ga, or Al, on the basis of our
all-electron, charge- and spin-selfconsistent KKR-CPA computations.
Specific compositions calculated include the concentrated
alloys with 25, 50, and 75 atomic percent V substituted
randomly in the Fe(B) site, the end compounds Fe$_3$X and Fe$_2$VX,
and the limiting cases of a single V impurity in
Fe$_3$X and a single Fe(B) impurity in Fe$_2$VX.
All calculations were carried to a high degree of
selfconsistency and are parameter-free excepting the use of experimental
lattice constants. The Fermi energy was evaluated in all cases with a high
degree of precision by using a generalized Lloyd formula for multi-component
systems. The use of a tetrahedron-type {\bf k} space integration method,
which we have extended to handle disordered muffin-tin systems,
allows us to treat the ordered end compounds as well as the single impurity
limits on an equal footing with the concentrated alloys.
Our results thus provide a reliable
basis for testing the underlying framework of the
KKR-CPA and LSD approximations. To our knowledge, little theoretical
work exists in the literature concerning the electronic structure of
disordered phases of Heusler alloys,
although Ref.~\onlinecite{kudrnovsky91} has previously
considered aspects of Fe$_{3-x}$V$_{x}$Si based on non-selfconsistent
computations.

Highlights of our results are as follows. An examination of the
spin-dependent component densities of states (CDOS's) at various sites
in Fe$_{3-x}$V$_{x}$Si shows that the replacement of Fe(B) by V
                                              induces substantial
shifts in spectral weights, especially in the up-spin
Fe(A,C) CDOS; the down-spin Fe(A,C) CDOS and the CDOS
for either spin direction at other sites -- Fe(B) or Si --
suffer relatively lesser changes. A pseudogap develops around
the Fermi energy in the up- as well as the down-spin
density of states in the V-rich regime. The effects of V substitution
in Fe$_{3-x}$V$_{x}$Ga are similar to those in Fe$_{3-x}$V$_{x}$Si,
although a number of differences arise from the reduced metalloid
valence and the changes in the metal-metalloid interaction.
Fe$_{3-x}$V$_{x}$Al, on the other hand, possesses a spectrum rather close
to that of Fe$_{3-x}$V$_{x}$Ga, some differences in detail notwithstanding.

The complex spectral changes in the electronic spectra of
Fe$_{3-x}$V$_{x}$X induced by Fe(B)/V substitution and by the effects
of metal-metalloid interaction cannot be described within a rigid band
picture based on one of the end compounds. In a similar way, the
ferromagnetism of these alloys cannot be viewed as a rigid splitting
of the related paramagnetic spectra. However, the common band model in
which the bonding between atoms is described by the formation of a
common energy band containing bonding and antibonding states with
different weights on the participating atoms
provides insight into the electronic structure of these
compounds. 

We have analyzed the composition dependence of the magnetic moments
associated with individual sites in Fe$_{3-x}$V$_{x}$X in detail,
and correlated these changes with those in the underlying
spectra. In Fe$_{3-x}$V$_{x}$Si, all moments vary essentially linearly
for $x \le 0.50$, and remain virtually constant thereafter.
V substitution (up to 50$\%$ V) decreases the Fe(A,C) moment, but
{\em increases} the Fe(B) moment;
Si possesses a very small (-0.08 $\mu_{B}$) negative moment.
The V impurity in Fe$_3$Si possesses a negative moment of
-0.62 $\mu_{B}$. The total moment in Fe$_{3-x}$V$_{x}$Si
varies non-linearly over $0\le x\le 1.0$ with a change in slope at
$x=0.50$. This behavior differs sharply from that expected on the basis
of the `local environment' picture which assumes the Fe(A,C) moment
to decrease linearly over $0\le x\le 1.0$ and the Fe(B) moment to
remain constant. On the other hand, the moments in Fe$_{3-x}$V$_{x}$Ga
and Fe$_{3-x}$V$_{x}$Al show a different type of non-linearity
in that the Fe(A,C) moment decreases rapidly between $0.5 \le x \le 0.75$,
and the aforementioned `saturation' effect in Fe$_{3-x}$V$_{x}$Si
for $x > 0.50$ is not seen. These results show clearly that the
metal-metalloid interaction has an important bearing on the
magnetism of Heusler alloys. Some measurements of the total magnetic
moment in Fe$_{3-x}$V$_{x}$Si, Fe$_{3-x}$V$_{x}$Ga, and
Fe$_{3-x}$V$_{x}$Al are available, and in this regard our theoretical
results are in reasonable agreement with the experimental values.
Further experimental work particularly to obtain site decomposed
moments in Fe$_{3-x}$V$_{x}$X should prove worthwhile.
Finally,      we have delineated the nature of carriers involved in
transport processes in Fe$_{3-x}$V$_{x}$X.
In the $x=0$ limit, most of the carriers originate
in the down-spin Fe(A,C) sites. With increasing
V concentration, as a pseudogap forms around the Fermi energy,
the total number of carriers in Fe$_{3-x}$V$_{x}$Si decreases rapidly.
In the V-rich alloy, the number of carriers begins to increase once
again, but it is now dominated by up-spin electrons.
By contrast, the V-rich compounds in the case of Fe$_{3-x}$V$_{x}$Ga
or Fe$_{3-x}$V$_{x}$Al are predicted
to be semimetallic, and to be dominated by down-spin carriers throughout
the composition range. These dramatic changes in the number, spin,
and type of carriers in Fe$_{3-x}$V$_{x}$X and their interaction with
strong local moments of Fe(B) impurities may play an important role in the
anomalous behavior of the resistivity and other transport properties
and in the possible heavy-fermion character of some of these compounds.

%%%%
%--------------------------------------------------------------------------
\acknowledgments
This work is supported by the US Department of Energy under contract
W-31-109-ENG-38, including a subcontract to Northeastern University,
the Polish Council of Science and Research
through grant number 2-P302-10307, and the allocation of supercomputer
time at the NERSC and the 
Northeastern University Advanced Scientific Computation Center (NU-ASCC). 
We thank NATO for a travel grant.

%%%%
%--------------------------------------------------------------------------

%%%captions
%--------------------------------------------------------------------------
%FIG. 1
\begin{figure}
\caption{
Crystal structure of Fe$_{3-x}$V$_{x}$X. The four crystallographic positions,
denoted by A-D, are shown by shadings of different intensity.
The A and C sites are equivalent in the present case.
Fe and V atoms occupy B sites,
while Si, Ga, or Al atoms sit at D sites.
}
\label{fig1}
\end{figure}
%FIG. 2
\begin{figure}
\caption{
Schematic diagram of the up- and-down spin component densities of states
on Fe(A,C), Fe(B), and V(B) sites
for a) Fe$_3$Si and b) Fe$_{2}$VSi. The horizontal dashed
lines mark the Fermi levels for Fe$_3$Si and Fe$_2$VSi, the chain lines
for the corresponding Ga and Al compounds.
The kinks in the latter represent an overall 
shift of the entire set of down-spin bands. The tick marks on the
energy axes denoted Fe(A,C), Fe(B), V, and Si(p) give
the various on-site metal $d$ and metalloid $p$ energy
levels. Tick marks denoted V in a) refer to the energy levels of a single
V impurity in Fe$_3$X; similarly Fe(B) in b) refers to single Fe(B) 
impurity levels in Fe$_2$VX.
}
\label{fig2}
\end{figure}
%FIG. 3
\begin{figure}
\caption{
Schematic representation of the common-band model of bonding in a
binary AB system. a) Densities of states of solids A and B before alloying;
b) component densities of states in the alloy AB; c) total density of states
in the alloy AB.
W denotes the width of the common band, W$_{AB}$ the
bandwidth after bonding. The quantities $E_A^0$ and $E_B^0$ give the
free-atom energy levels while $E_A$ and $E_B$ denote the local on-site
energy levels. After Ref.\protect\onlinecite{pettifor95}.
}
\label{fig3}
\end{figure}
%FIG. 4
\begin{figure}
\caption{
Computed total magnetic moment (per Wigner-Seitz cell) and the moments per
atom on various inequivalent sites in Fe$_{3-x}$V$_{x}$Si as a function
of V concentration $x$. All values given in units of Bohr magnetons ($\mu_B$).
Different symbols are explained in the legend. Lines are drawn through the
theoretical points to guide the eye. Experimental points (open circles)
for the total moment are after Ref.\protect\onlinecite{niculescu83}.
}
\label{fig4}
\end{figure}
%FIG. 5
\begin{figure}
\caption{
Same as Fig.\protect\ref{fig4}, except that this figure refers to
Fe$_{3-x}$V$_{x}$Ga. Experimental points after
Ref.\protect\onlinecite{kawamiya83}.
}
\label{fig5}
\end{figure}
%FIG. 6
\begin{figure}
\caption{
Same as Fig.\protect\ref{fig4}, except that this figure refers to
Fe$_{3-x}$V$_{x}$Al. Experimental points after
Refs.\protect\onlinecite{popiel89} and
\protect\onlinecite{tuszynski90}.
}
\label{fig6}
\end{figure}
%FIG. 7
\begin{figure}
\caption{
Component density of states (CDOS) for various inequivalent sites in Fe$_3$Si
(topmost row). Different angular momentum contributions to the
CDOS are shown. The majority (up) and
minority (down) spin part of the CDOS is given in each case.
The dotted vertical lines mark the Fermi energy ($E_F$).
Note different scales.
}
\label{fig7}
\end{figure}
%FIG. 8
\begin{figure}
\caption{
Computed partial density of states in the $s$ and $d$ channels
associated with a Si site in Fe$_3$Si; the spectrum has been convoluted
with a Gaussian of 2.0 eV (full-width-at-half-maximum) to mimic
experimental broadening of the soft x-ray
emission spectra of Ref.\protect\onlinecite{jia92}.
}
\label{fig8}
\end{figure}
%FIG. 9
\begin{figure}
\caption{
Total and site-decomposed density of states in Fe$_3$Si,
Fe$_3$Ga, and Fe$_3$Al. See caption to Fig.\protect\ref{fig7}
for other details.
}
\label{fig9}
\end{figure}
%FIG. 10
\begin{figure}
\caption{
Same as Fig.\protect\ref{fig9}, except that this figure refers to
Fe$_2$VX alloys.
}
\label{fig10}
\end{figure}
%FIG. 11
\begin{figure}
\caption{
Dispersion curves along high symmetry lines in the Brillouin zone
in ferromagnetic Fe$_{2}$VSi (upper) and paramagnetic Fe$_{2}$VGa (lower).
The darker and lighter curves in the upper picture represent spin-up and
spin-down bands in Fe$_{2}$VSi respectively. The Fermi energy is marked by
horizontal lines.
}
\label{fig11}
\end{figure}
%FIG. 12
\begin{figure}
\caption{
Site-dependent density of states
in Fe$_{3-x}$V$_{x}$Si at Fe(A,C), Fe(B),
and V(B) sites for each spin direction as a function
of the V concentration $x$.
The Fermi energy is marked by dotted vertical lines.
}
\label{fig12}
\end{figure}
%FIG. 13
\begin{figure}
\caption{
Total up- and down-spin density of states
in Fe$_{3-x}$V$_{x}$Si, Fe$_{3-x}$V$_{x}$Ga,
and Fe$_{3-x}$V$_{x}$Al for different V concentrations $x$.
The Fermi energies are marked by dotted vertical lines.
}
\label{fig13}
\end{figure}
%FIG. 14
\begin{figure}
\caption{
Same as Fig.\protect\ref{fig12}, except that this figure refers to
Fe$_{3-x}$V$_{x}$Ga.
}
\label{fig14}
\end{figure}
%FIG. 15
\begin{figure}
\caption{
The differences between the total energies of Fe$_{3-x}$V$_{x}$Si
and constituent atoms with V in the B sublattice (solid curve)
and V in the A or C sublattices (dashed curve).
}
\label{fig15}
\end{figure}
%--------------------------------------------------------------------------
%%%1

%%%tables
\begin{table}
\caption{
 The lattice parameters ($a$) and the muffin-tin radii
($S_k$) in Fe$_3$X.
}
\begin{tabular}{lll}
 & $a$(\AA) & $S_k$(\AA) \\
 \cline{1-3}
Fe$_3$Si& 5.653\tablenotemark[1]& 1.224 \\
Fe$_3$Ga& 5.812\tablenotemark[2]& 1.258 \\
Fe$_3$Al& 5.791\tablenotemark[3]& 1.254 \\
\end{tabular}
\tablenotetext[1]{Ref.\ \onlinecite{niculescu76}.}
\tablenotetext[2]{Ref.\ \onlinecite{kawamiya82}.}
\tablenotetext[3]{Ref.\ \onlinecite{motoya83}.}
\label{table1}
\end{table}

\begin{table}
\caption{
Magnetic moments on various inequivalent sites in the end
compounds Fe$_3$X and Fe$_2$VX in units of Bohr magnetons ($\mu_B$).
The results of the present computations are given together with
those of a number of other computations and relevant experiments.
The values of moments for a single V(B)
impurity in Fe$_3$X and a single Fe(B) impurity in Fe$_2$VX
are based on our KKR-CPA computations in the limiting compounds,
and are marked with stars.
}
\begin{tabular}{c|c|c|c|c|c|c|c|c|c|c|c}
 \multicolumn{1}{c|}{ } &
 \multicolumn{3}{c|}{Fe(A,C)} &
 \multicolumn{3}{c|}{Fe(B)} &
 \multicolumn{1}{c|}{V(B)} &
 \multicolumn{2}{c|}{X(D)} &
 \multicolumn{2}{c }{Total } \\
 \cline{2-12}
 \multicolumn{1}{c|}{ } &
 \multicolumn{2}{c|}{Theory}&
 \multicolumn{1}{c|}{Exp.} &
 \multicolumn{2}{c|}{Theory} &
 \multicolumn{1}{c|}{Exp.} &
 \multicolumn{1}{c|}{Theory} &
 \multicolumn{2}{c|}{Theory} &
 \multicolumn{2}{c }{Theory} \\
 \multicolumn{1}{ c|}{ } &
 \multicolumn{1}{ c|}{present} &
 \multicolumn{1}{ c|}{others} &
 \multicolumn{1}{ c|}{ } &
 \multicolumn{1}{ c|}{present} &
 \multicolumn{1}{ c|}{others} &
 \multicolumn{1}{ c|}{ } &
 \multicolumn{1}{ c|}{present} &
 \multicolumn{1}{ c|}{present} &
 \multicolumn{1}{ c|}{others} &
 \multicolumn{1}{ c|}{present}&
 \multicolumn{1}{ c }{others} \\
\cline{1-12}
 Fe$_3$Si &  1.34 &   1.2\tablenotemark[1] & 1.2\tablenotemark[2]
          &  2.52 &   1.9\tablenotemark[1] & 2.4\tablenotemark[2]
          & -0.62*
          & -0.08 &
          &  5.07 &
\\
          &       &  1.10\tablenotemark[3] & 1.07\tablenotemark[4]
          &       &  2.52\tablenotemark[3] & 2.23\tablenotemark[4]
          &
          &       & -0.08\tablenotemark[4]
          &       &  4.63\tablenotemark[4]
\\
          &       &                        & 1.35\tablenotemark[5]
          &       &                        & 2.20\tablenotemark[5]
          &
          &       &
          &       &
\\
\cline{1-12}
Fe$_3$Ga  &  1.87 & 2.09\tablenotemark[6]  &  1.7\tablenotemark[7]
          &  2.47 & 2.38\tablenotemark[6]  &  2.5\tablenotemark[7]
          & -1.16*
          & -0.08 &-0.10\tablenotemark[6]
          &  6.01 &
\\
\cline{1-12}
Fe$_3$Al  &  1.90 & 1.36\tablenotemark[8]  & 1.5 \tablenotemark[2]
          &  2.44 & 2.48\tablenotemark[8]  & 2.18\tablenotemark[2]
          & -1.27*
          & -0.10 &
          &  6.03 &
\\
          &       & 1.26\tablenotemark[9]  &
          &       & 2.16\tablenotemark[9]  &
          &
          &       &
          &       &
\\
          &       & 1.89\tablenotemark[10]  &
          &       & 2.34\tablenotemark[10]  &
          &
          &       & -0.14\tablenotemark[10]
          &       &
\\
          &       & 1.9\tablenotemark[11]  &
          &       & 2.4\tablenotemark[11]  &
          &
          &       &
          &       &
\\
\cline{1-12}
Fe$_2$VSi &  0.43 &  0.37\tablenotemark[3] &
          &  3.08*&                        &
          & -0.11
          & -0.02 & -0.02\tablenotemark[3]
          &  0.68 &  0.70\tablenotemark[3]
\\
\cline{1-12}
Fe$_2$VGa &  0.03 &                        &
          &  3.20*&                        &
          & -0.06
          &  0.00 &
          &  0.00 &
\\
\cline{1-12}
Fe$_2$VAl &  0.03 &                        &
          &  3.18*&                        &
          & -0.04
          &  0.00 &
          &  0.00 &
\\
\end{tabular}
\tablenotetext[1]{Ref.\ \onlinecite{garba86}.}
\tablenotetext[2]{Ref.\ \onlinecite{pickart61,paoletti64}.}
\tablenotetext[3]{Ref.\ \onlinecite{kudrnovsky91}.}
\tablenotetext[4]{Ref.\ \onlinecite{moss72}.}
\tablenotetext[5]{Ref.\ \onlinecite{hines76}.}
\tablenotetext[6]{Ref.\ \onlinecite{ishida89}.}
\tablenotetext[7]{Ref.\ \onlinecite{kawamiya82}.}
\tablenotetext[8]{Ref.\ \onlinecite{williams82}.}
\tablenotetext[9]{Ref.\ \onlinecite{ishida76}.}
\tablenotetext[10]{Ref.\ \onlinecite{guo98}.}
\tablenotetext[11]{Ref.\ \onlinecite{singh98}.}
\label{table2}
\end{table}

%--------------------------------------------------------------------------

\begin{references}
\bibitem{footnote3}
A discussion of cases where more than one sublattice is disordered
simultaneously is considered outside the scope of this work.
\bibitem{footnote4}
 For properties considered in this article, only the `diagonal'
 matrix elements of the Green function,
 where one stays within the central unit cell, are required.
 More general `off-diagonal' matrix elements needed for computing,
for example, momentum density, (cf. Refs.~\onlinecite{mijnarends76}
and \onlinecite{bansil81}) are not considered in
Eq.~(\ref{eqover:01}).
\bibitem{footnote6}
  The use of $\ell_{max}=2$ on Si sites is thus important.
\bibitem{footnote7} For example, comparing the $x=0.75$ and $x=0.0$
                    Fe(B) results (Figs.~\ref{fig12}b4 and
                    \ref{fig12}b1) we see that the up-spin CDOS is
                    much smoother for $x=0.75$ even though this is
                    hardly so for the down spins.
\bibitem{footnote5} However, care should be excercised in this respect
                    since it is well-known that the local-density
                    approximation (LDA) tends to underestimate the
                    band gap in semiconductors.
\end{references}
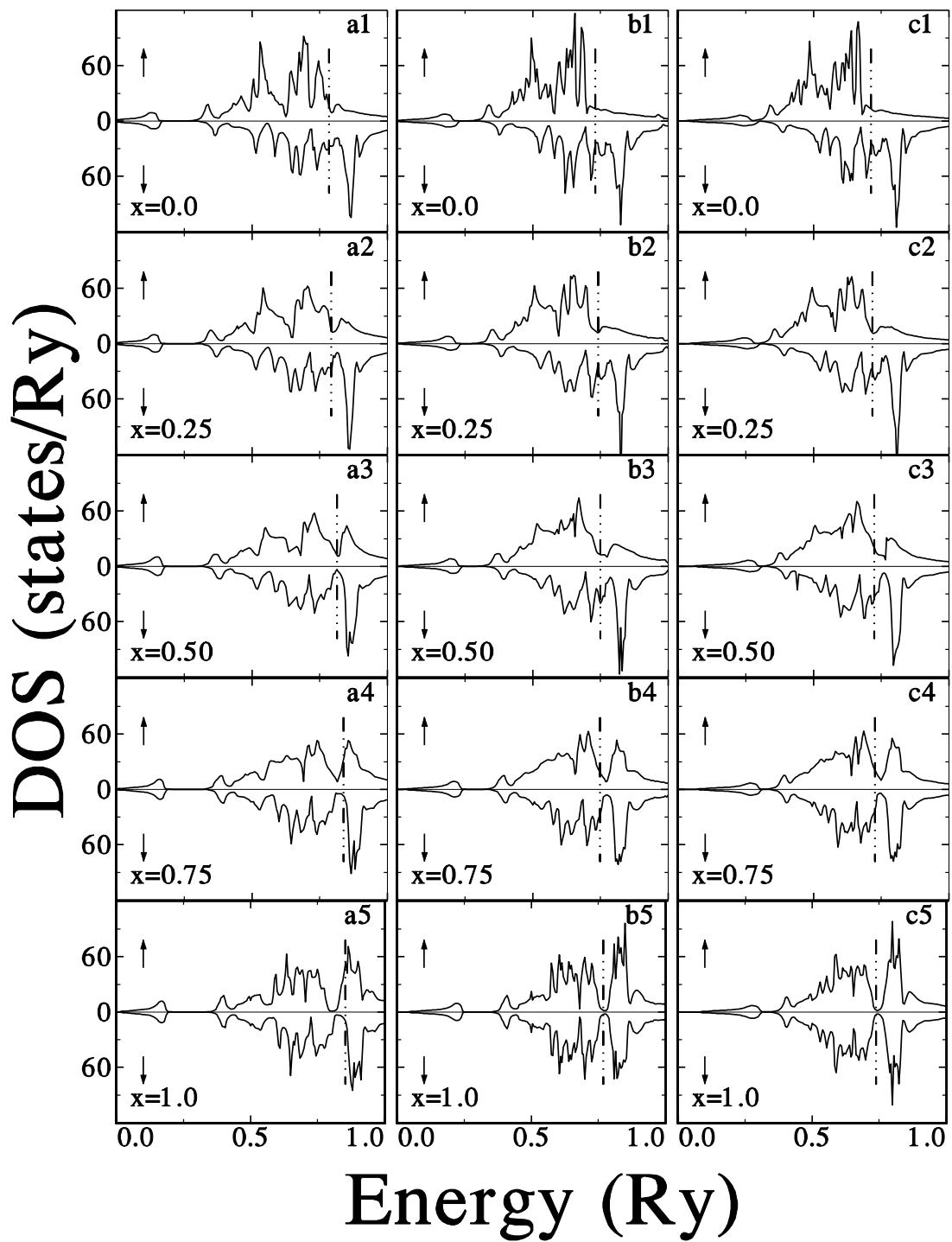
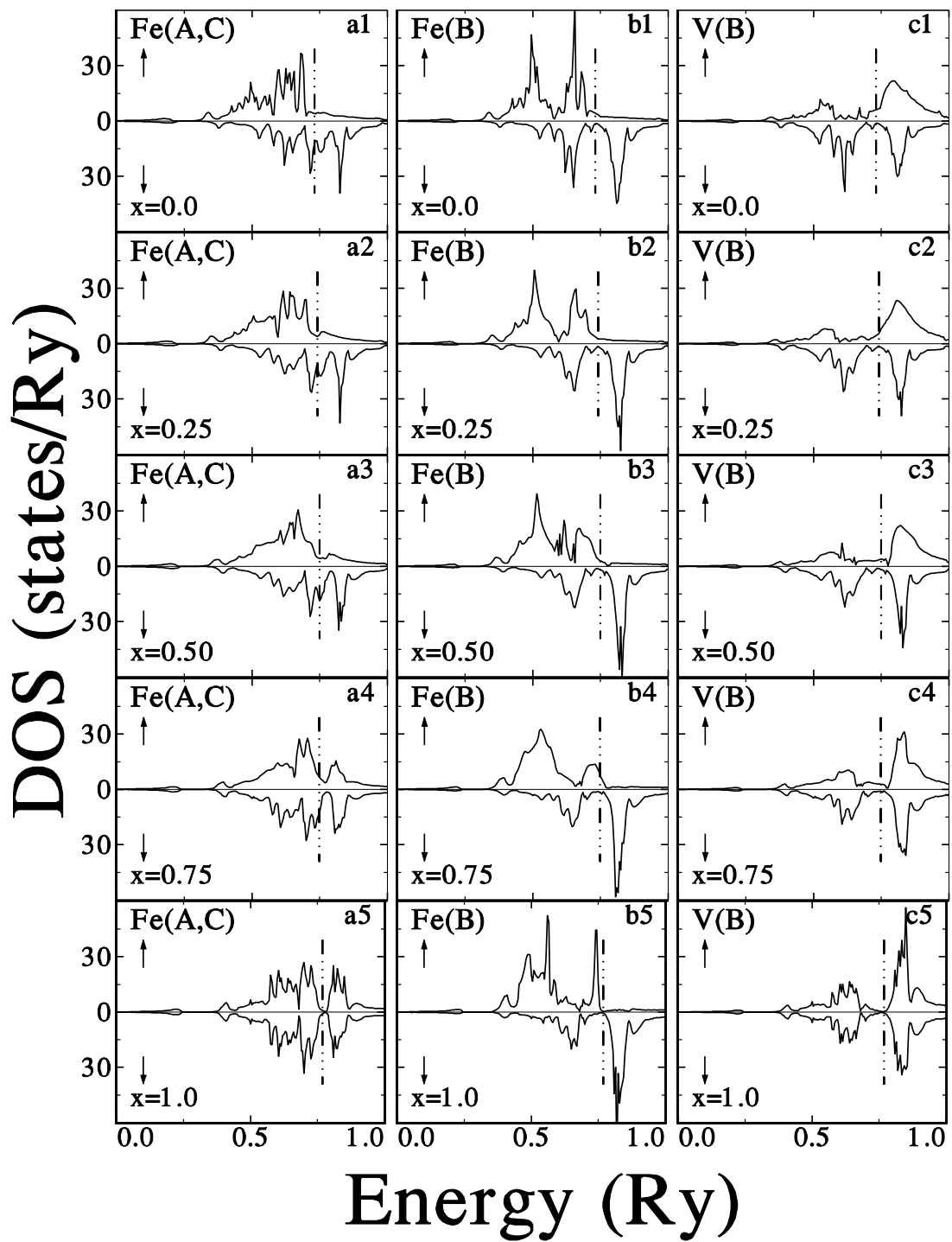
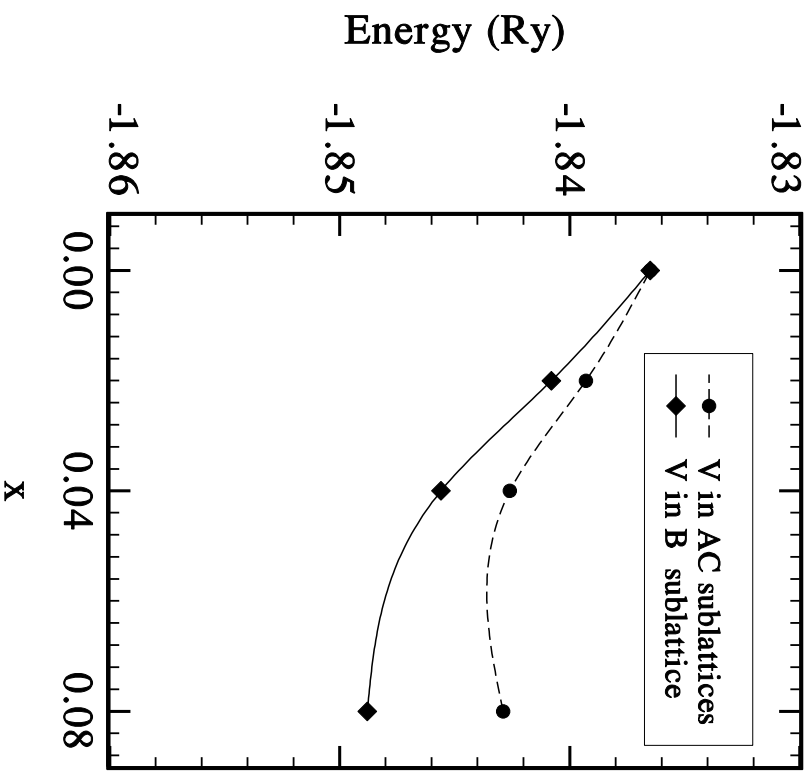
\end{document}